\documentclass[useAMS,usenatbib]{mn2e}
\usepackage{graphicx}

\def\mnras{MNRAS}
\def\apj{ApJ}
\def\apjl{ApJ}

\def\aj{AJ}
\def\pasp{PASP}
\def\aap{A\&A}
\def\aaps{A\&AS}  
\def\kms{km\,s$^{-1}$}
\def\ene{erg\,s$^{-1}$}

\def\Ha{H$\alpha$}
\def\Hb{H$\beta$}

\def\LL{$\lambda\lambda$}
\def\L{$\lambda$}
\def\CaII{Ca\,{\sc ii}}
\def\CII{C\,{\sc ii}} 
\def\CI{C\,{\sc i}} 
\def\OI{O\,{\sc i}}
\def\OII{O\,{\sc ii}}
\def\OIII{O\,{\sc iii}}

\def\SiII{Si\,{\sc ii}}
\def\TiII{Ti\,{\sc ii}}
\def\SII{S\,{\sc ii}}
\def\ScII{Sc\,{\sc ii}}
\def\HeI{He\,{\sc i}}
\def\FeII{Fe\,{\sc ii}}

\def\MgII{Mg\,{\sc ii}}
\def\MgI{Mg\,{\sc i}}
\def\NaID{Na\,{\sc i}~D}
\def\NaI{Na\,{\sc i}}
\def\NeI{Ne\,{\sc i}}
\def\NII{N\,{\sc ii}}
\def\to{$\rightarrow$}
\def\msun{M$_{\odot}$}

\title[SN 2009jf]{SN 2009jf:  a slow-evolving stripped-envelope core-collapse supernova\thanks{This paper is based on observations with several telescopes, including: NTT(184.D-1151), VLT-UT1(085.D-0750,386.D-0126), NOT, WHT, TNG, PROMPT, Ekar, Calar Alto, Liverpool Telescope.}} 

\author[Valenti et al.]{S. Valenti$^{1,2}$
\thanks{e--mail: stefano.valenti@oapd.inaf.it},
M. Fraser$^{1}$, S. Benetti$^{2}$, G. Pignata$^{3,4}$, J. Sollerman$^{5}$, C. Inserra $^{6,7}$,
\and
 E. Cappellaro$^{2}$ , A. Pastorello$^{1}$, S. J. Smartt$^{1}$, M. Ergon$^{5}$, M. T. Botticella$^{2}$,
\and
J. Brimacombe$^{8}$, F. Bufano$^{2}$, M. Crockett$^{9}$, I. Eder$^{10}$,  D. Fugazza$^{11}$,  J. B Haislip$^{12}$, 
\and
M. Hamuy$^{4}$,  A. Harutyunyan$^{13}$,K. M. Ivarsen$^{12}$, E. Kankare$^{14,15}$, R. Kotak$^{1}$,    
\and
A. P. LaCluyze$^{12}$, L. Magill$^{1}$, S. Mattila$^{5,14}$,  J. Maza$^{4}$, P. A. Mazzali$^{16,2}$,
\and
D. E. Reichart$^{12}$, S. Taubenberger$^{16}$,  M. Turatto$^{17}$, L. Zampieri$^{2}$,
\and
 \\ 
$^{1 ~}$ Astrophysics Research Centre, School of Mathematics and Physics,
Queen's University Belfast, Belfast BT7 1NN, United Kingdom\\
$^{2 ~}$  INAF Osservatorio Astronomico di Padova, Vicolo dell'Osservatorio 5, 35122 Padova, Italy \\
$^{3 ~}$ Departamento de Ciencias Fisicas, Universidad Andres Bello, Avda. Republica 252, Santiago, Chile \\
$^{4 ~}$ Departamento de Astronomia, Universidad de Chile, Casilla 36-D, Santiago, Chile\\ 
$^{5 ~}$ The Oskar Klein Centre, Department of Astronomy, Stockholm University, AlbaNova, 10691 Stockholm, Sweden \\
$^{6 ~}$ INAF Osservatorio Astrofisico di Catania, Via S. Sofia 78, 95123, Catania, Italy\\
$^{7 ~}$ Dipartimento di Fisica ed Astronomia, Universit\'a di Catania, Sezione Astrofisica, Via S.Sofia 78, 95123, Catania, Italy\\
$^{8 ~}$ Coral Towers Observatory, Coral Towers, Esplanade, Cairns 4870, Australia. \\
$^{9 ~}$ Department of Physics, University of Oxford, Keble Road, Oxford OX1 3RH\\
$^{10}$ Medve u. 15, H-1027 Budapest, Hungary  \\
$^{11}$ INAF-Osservatorio Astronomico di Brera, via Bianchi 46, 23807 Merate, Italy \\
$^{12}$ University of North Carolina at Chapel Hill, Campus Box 3255, Chapel Hill, NC 27599-3255, USA\\
$^{13}$ Fundaci\'on Galileo Galilei-INAF, Telescopio Nazionale Galileo, 38700 Santa Cruz de la Palma, Tenerife, Spain\\
$^{14}$ Tuorla Observatory, Department of Physics $\&$ Astronomy, University of Turku, V\"{a}is\"{a}l\"{a}ntie 20, FI-21500 Piikki\"{o}, Finland \\
$^{15}$ Nordic Optical Telescope, Apartado 474, E-38700 Santa Cruz de La Palma, Spain \\
$^{16}$ Max-Planck-Institut f\"{u}r Astrophysik, Karl-Schwarzschild-Str. 1, 85741 Garching bei M\"{u}nchen, Germany\\
$^{17}$ INAF Osservatorio Astronomico di Trieste, Via Tiepolo 11, 34143, Trieste, Italy \\
}

\begin{document}

\date{Accepted 15/06/2011; Received 12/05/2011;}

\pubyear{2011}

\maketitle

\begin{abstract}

We present an extensive set of photometric and spectroscopic data for
SN 2009jf, a nearby Type Ib supernova, spanning from $\sim$ 20 days
before $B$-band maximum to one year after maximum.  We show that SN
2009jf is a slowly evolving and energetic stripped-envelope SN and is likely 
from a massive progenitor (25-30 solar masses).
The large progenitor's mass allows us to explain the complete hydrogen plus 
helium stripping without invoking the presence of a binary companion. 
The supernova occurred close to a young cluster, in a crowded environment 
with ongoing star-formation.  
The specroscopic similarity with the He-poor Type Ic SN 2007gr
suggests a common progenitor for some supernovae Ib and Ic.  The
nebular spectra of SN 2009jf are consistent with an asymmetric
explosion, with an off-center dense core.  We also find evidence that
He-rich Ib supernovae have a rise time longer than other
stripped-envelope supernovae, however confirmation of this result and
further observations are needed.
\end{abstract}


\section{Introduction}
\label{parintroduction}
Type Ib supernovae (SNe) are observationally defined by the absence of
hydrogen (H) and presence of helium (He) in their spectra at early
phases. They are thought to come from massive stars (M $\ge$ 8
M$_{\odot}$) which have lost their H envelope, but retained part of
their He envelope before exploding. SNe Ic show neither H nor He
features in their early spectra, and arise from massive stars which
have also lost most of their He envelope. A group of SNe that show H
features at early phases, but do not follow the typical light curve
seen for the H-rich Type II SNe, and instead evolve photometrically
and spectroscopically into SNe Ib are known as Type IIb SNe.  These
are also thought to come from massive star progenitors which have lost
most (but not all) of their H envelope before exploding.  SNe IIb, Ib
and Ic are often collectively termed ``stripped-envelope SNe''
\citep{clocchiatti97}.

We emphasize that the apparent lack of H and/or He lines in an early
spectrum does not preclude the presence of these elements in the
ejecta. \cite{branch02} analysed a large sample of SN Ib spectra, and
suggested that H is often present also in SNe Ib, though this element
is very difficult to detect after maximum. They also found that He is
often present outside the photosphere (detached), in particular at
late phases.  Similar studies of SNe Ic \citep{branch06,elmhamdi06}
suggest that traces of H and He may be also present in some SNe Ic.
The presence of He and/or H in some stripped-envelope SNe could be an
indication that the progenitor has been stripped continuously (which
can leave thin layers of H or He on the progenitor) as opposed to
strong episodic mass loss, which would likely completely strip the
progenitor.  Furthermore, the small observational differences between
He rich (Ib) and He poor (Ic) SNe may suggest a similar origin, with
progenitors of SNe Ic simply being more strongly stripped than those
of SNe Ib.

An important open issue on SNe Ib/c is whether they come from
relatively high mass Wolf-Rayet (WR) stars (M $> 20-25$ M$_{\odot}$),
that have been stripped of H and part of their He envelope by
radiatively-driven winds, or from lower mass progenitors (M $>$ 11
M$_{\odot}$) which have been stripped of their envelope by a binary
companion \citep{Podsiadlowski92}.

Several statistical studies have been performed on the environments
of stripped-envelope SNe in order to characterise the Ib and Ic
progenitor populations. Many have focused on metallicity as a key
parameter driving mass loss in single stars and hence determining the
relative numbers of Types II, Ib and Ic SNe. Metallicity is also
important in binary models \citep{Eldridge08}, and influences the
number of WR stars produced as well as the predicted ratios of Ibc SN.
Rotation rates are also thought to be dependent on initial stellar 
metallicity \citep{Georgy09}.
Initial studies of the global properties of host galaxies suggested
that the number of SNe Ic relative to the number of
 SNe Ib increases with metallicity  \citep{prieto08,Boissier09,Anderson09}.
However more recent measurements of the nebular 
oxygen abundances at  the sites of SNe Ib and Ic  now 
suggest there is little difference between the samples
\citep{Modjaz11,Anderson10,Leloudas11}.
From these results there is no unambiguous evidence that 
Ibc SNe are produced in more metal rich regions than type II SNe, or 
 that Ib and Ic SNe originate in different metallicity regimes.

The relatively high frequency of SNe Ib/c (30 per cent of all core-collapse SNe by volume)
and the lack of progenitor detections \citep{Crockett09,smartt09b}
supports the idea that at least some
of the SNe Ib/c progenitors are massive stars in binary systems
\citep{Podsiadlowski92,fryer07,Eldridge08}.  On the other hand, the
extreme kinetic energies inferred for some stripped-envelope SNe,
namely those associated with GRBs, are indicative of a high-mass
progenitor which does not need necessary binarity to lose its envelope.

Excluding the extreme cases, for most stripped-envelope SNe
discriminating between the progenitor scenarios is more tricky.  In
principle one may argue that SNe which eject more material probably
come from more massive stars, while SNe with low mass ejecta may come
from less massive progenitors in binary systems. On the other hand, if
fall-back occurs (where a portion of the SN ejecta falls back on to
the newly born compact remnant at the centre) even very massive stars
may eject only a small amount of material. A detailed analysis of the
elemental abundances in the ejecta is then a useful diagnostic.  For
example, oxygen and carbon are expected to be more abundant in ejecta
of a SN from a massive progenitor, due to the larger progenitor core
mass. However, from an observational point of view, the number of
stripped-envelope SNe with detailed ejecta characterization is still
small. Recent discoveries have also pointed out the largely unexplored
diversity in He-rich SNe, for example the wide range of ejected mass
and energy as typified by SN 2007Y \citep{stritzinger09} and SN 2008D
\citep{mazzali08}, respectively, or the case of SN 1999dn, for years
considered a prototype of SNe Ib, but recently proposed to be a highly
energetic SN from a massive progenitor \citep{benetti11}.

In this framework, every new nearby SN Ib discovered represents an
opportunity to increase our understanding of these events.  SN 2009jf,
a nearby SN which was discovered early after the explosion, is a
perfect target to enlarge the sample of well-studied SNe Ib.

\begin{figure}
   \includegraphics[width=8.5cm,height=8.5cm]{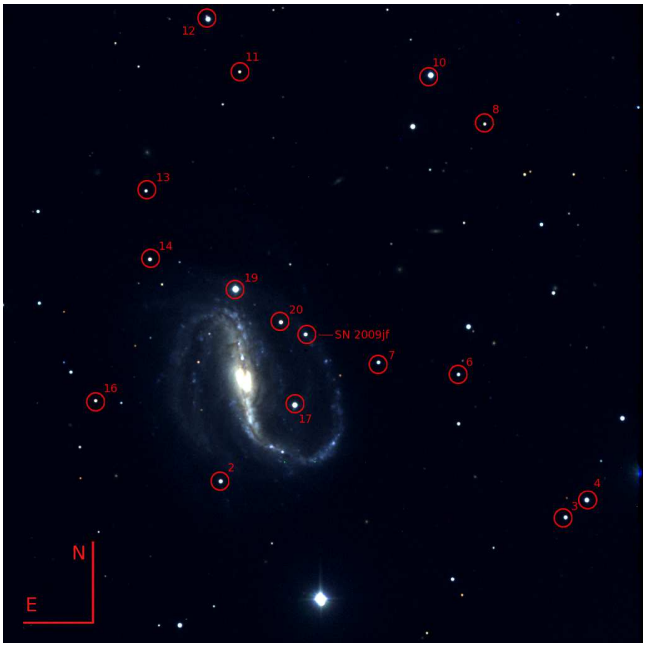}
  \caption{The field of view (9'x9') of SN 2009jf. This $R$-band Calar 
  Alto image (18 November 2009) was taken 34 days after the $B$-band maximum.
  The local sequence stars have been numbered to calibrate the photometry 
  (magnitudes reported in Tables \ref{tabseqstar1} and \ref{tabseqstar2}).}
  \label{figseqstar}
\end{figure}

In this paper we present the full set of data we collected for SN
2009jf, from the ultraviolet (UV) to optical and near infrared
(NIR). The structure of the paper is as follows: in the next section
we describe the set of data collected and its calibration.  In
Section~\ref{parphot} we present the photometric data and in
Sections~\ref{parspe} and \ref{spenebular} the optical and NIR spectra
of SN 2009jf.  In Section~\ref{sec:host}, we discuss the properties
of the host galaxy NGC 7479 (see Fig. \ref{figseqstar}) and provide an
analysis of the progenitor using archival pre-explosion data. In
Section~\ref{parametribolo}, we discuss the main physical parameters
of the progenitor of SN 2009jf at the moment of the explosion through
an analysis of the bolometric light curve. In the final section, we
summarise our results and discuss our conclusions.  We note that
\cite{Sahu11} have also presented a spectro-photometric study of SN
2009jf in the optical with which we will compare our results with.

\section{Discovery and follow-up}
\label{parobs}

SN 2009jf was discovered \citep{li2009} on 2009 September 27.33 (UT
dates are used throughout this paper) with the Katzman Automatic
Imaging Telescope (KAIT) during the Lick Observatory Supernova Search
\citep{filippenko01}. The supernova is located at coordinates
$\alpha~$ = $23^h 04^m 52^s.98$ and $\delta$ =$ +12\degr 19' 59''.5$
(equinox J2000), which is $53''.8$ W and $36''.5$ N of the centre of
the host galaxy NGC 7479. The host is a barred spiral galaxy, with an
intriguing jet-like radio continuum feature. The alignment of this
jet, which is in the opposite orientation to the optical arms, has
been suggested to be consistent with NGC 7479 having recently
undergone a minor merger \citep{laine08}. SN 2009jf was not visible in
a KAIT unfiltered image taken 4 days before discovery (September
23.32) \citep[$>$19.2 mag,][]{li2009} and was classified on September
29.1 as a young Type Ib SN similar to SN 1999ex
\citep{Kasliwal09,sahu09}. \cite{Itagaki09} reported the detection of
a source close to the position of the SN in several images obtained over the
past few decades. A rough estimate of the absolute magnitude of the
source in the pre-discovery images ($-14.5$ mag) led \cite{Itagaki09}
to initially suggest a Luminous Blue Variable (LBV) as the progenitor
of SN 2009jf.  However, we have undertaken a more thorough analysis of
the archival images, and the source is more likely a cluster close to
the position where the SN occurred (see Section~\ref{sec:host}). SN
1990U, which was a SN Type Ic, also exploded in this galaxy
\citep[]{Pennypacker90,filippenko90b}.

Being discovered well before maximum, and in a nearby host galaxy, SN
2009jf was targeted for an intensive spectro-photometric
follow-up campaign by the European Large Programme (ELP) SN
Collaboration\footnote{http://graspa.oapd.inaf.it/index.php$?$option=com\_content\&view=article\&id=68\&Itemid=93}, 
together with the Millenium Center for Supernova Science (MCSS).

Our photometric and spectroscopic monitoring campaign for SN 2009jf
began on 2009 October 1st, just 7 days after explosion (see
Section~\ref{parphot}).  We observed the SN every $\sim2-3$ days in
Sloan and Bessel filters, with sligthly more relaxed coverage (one
observation every $\sim4-5$ days) in the NIR bands.  From the
beginning of December, $\sim$2.5 months after explosion, the SN was no
longer visible from the Southern Hemisphere. From then on, it was
observed from the Northern Hemisphere with a more relaxed cadence (one
observation every week) until it disappeared behind the sun at
$\sim$105 days after explosion. The SN was recovered as soon as it was
visible again in June 2010 with observations that extended until
October to cover the nebular phase.

We used several of the facilities available to the ELP collaboration,
and also the five PROMPT\footnote{Panchromatic Robotic Optical
  Monitoring and Polarimetry Telescopes.} \citep{Reichart05}
telescopes used by the MCSS project.  The \emph{Swift} telescope also
observed SN 2009jf at UV wavelengths, and the publicly available data
from this has been included in our analysis. However, due to the
strong contamination from the close-by cluster the \emph{Swift} $uvm2$
and $uvw2$ filter data are not usable (see Appendix~\ref{ap0}) and
thus not reported.

NGC 7479 is one of the most beautiful nearby face-on galaxies, and  a popular
target for amateur astronomers.  Some of the images obtained by 
amateurs have been useful in constraining the explosion epoch, and
these have been added to our dataset.
In particular, we obtained images of NGC 7479 taken on September 23,
24, 26 and 27, providing excellent coverage close to the explosion
epoch\footnote{http://eder.csillagaszat.hu/deepsky/350D/sn2009jf/sn2009jf$\_$eder$\_$en.htm}.

The $UBVRI$ data ranging from $\sim$1 to $\sim$380 days after
explosion are reported in Table~\ref{tablandolt}, while the $ugriz$
data are reported in Table~\ref{tabsloan}. All data calibrated to the
Landolt system are in the Vega system, while the data calibrated to
Sloan are in the AB system \citep{smith02}.

Spectroscopic monitoring started on 2009 October 1st, 7 days after
explosion and continued during the photospheric phase until the
beginning of the seasonal gap at $\sim$105 d after explosion.  More
spectra were collected in the nebular phase, when the SN became
visible again.  In total we collected 20 optical and 4 infrared
spectra of SN 2009jf (see Section \ref{parspe}).  Details of our data
reduction methods are reported in Appendix \ref{ap0}.

\subsection{Archival observations}

To search for a progenitor in pre-explosion data \citep[see ][ for a
  review]{smartt09b}, we queried all suitable publicly available image
archives of which we are aware.

The most useful images for constraining the pre-explosion environment
and progenitor of SN 2009jf are from the Wide-Field and Planetary
Camera 2 (WFPC2) on-board the {\it Hubble Space Telescope (HST)}. The
site of SN 2009jf was observed for a total of 2000 s in the $F569W$
filter and 1800 s in the $F814W$ filter on 1995 October 16.

Pre-explosion data were also available from many ground based observatories,
but these images, which have a typical seeing of ~1
arcsec, are not of sufficient quality to resolve the complex region
where the SN exploded (see Section \ref{parprogenitor}). The same is
true for the {\it Spitzer} IRAC and {\it XMM-Newton} OM images.

However, several of these deep pre-explosion ground based images were
used as templates for application of the template subtraction
technique for our photometry (see Sec. \ref{ap0}).  
In particular we used the following images: $U$, $B$ and
$I$ images from William Herschel Telescope (WHT) (1990 August 17, 17
and 18 respectively), a $V$-band image from the ESO New Technology
Telescope (NTT) (1992 December 2), and an $R$-band image from the
Nordic Optical Telescope (NOT) (2009 September 13).

To identify a potential progenitor in {\it HST} data we obtained deep images
of SN 2009jf and its environs on 2009 October 24 and 25 with the
ESO-VLT (Very Large Telescope) and NaCo (Nasmyth Adaptive Optics
System and Near-Infrared Imager and Spectrograph). We used the $K_\mathrm{S}$
filter with the S54 camera\footnote{pixel scale of 0.054\arcsec over a
  56\arcsec $\times$ 56\arcsec\ field of view.}.  A late time
observation was obtained on 2010 September 16, but the SN had faded in
the NIR to the point where it was not significantly brighter than the
nearby cluster, and so this data was of no use in locating the
progenitor.

\section{Photometry}
\label{parphot}

\begin{figure*}
   \includegraphics[width=16cm,height=8cm]{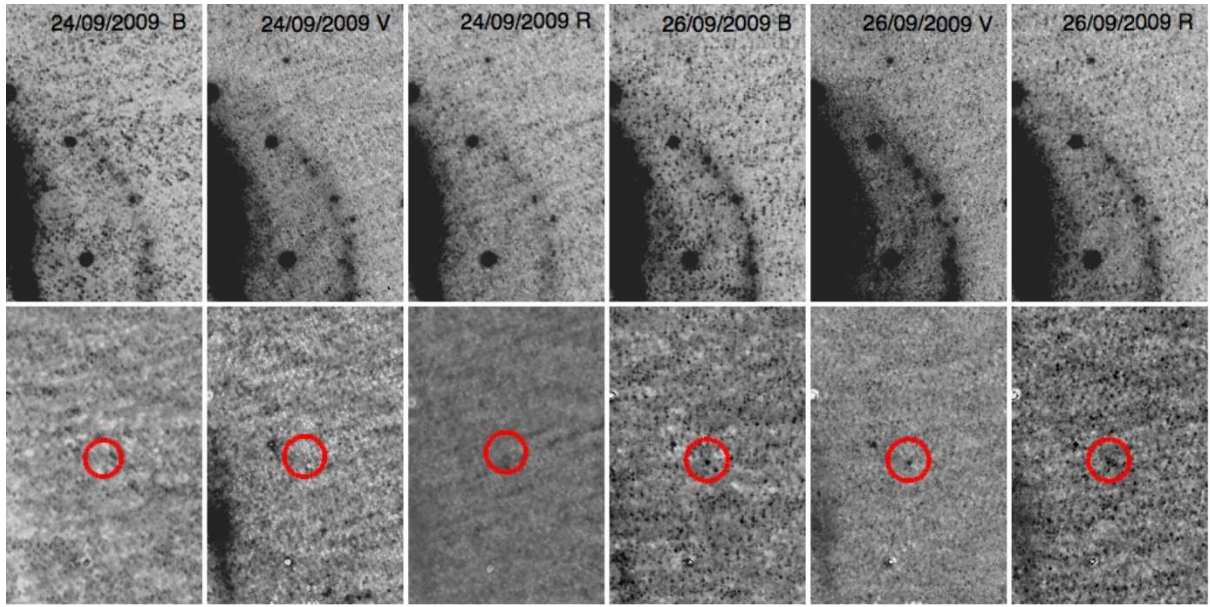}
  \caption{$BVR$ images of SN 2009jf obtained on 24th and 26th September 
  2009 (upper panels) together with difference images obtained by subtracting 
  images in the same filters obtained on 23rd September 2009 (lower panels). 
  The marginal detections on September 26th and the non-detections 
  on September 24th constrain the explosion epoch to 2009 
  September 25th (JD=2455099.5) $\pm 1$ which is one of 
  the best constrained explosion epochs for a stripped-envelope SN.
  North is up and East to the left.}
  \label{figexp}
\end{figure*}

To constrain the explosion epoch we used the pre-discovery images of
NGC 7479 taken by I. Eder on September 23rd--27th.  The
first observation on September 23rd was before the first non-detection
of SN 2009jf at the KAIT telescope \citep{li2009}.  Using these images
taken on September 23 as a template, we performed image subtraction on
the images taken over the following days (24th, 26th and 27th) with
the same telescope. The SN is marginally detected in $B$,$V$ and $R$ bands
 on September 26th, while nothing is detected
in the same bands on September 24th (see Fig.~\ref{figexp}). The
detections on September 26th and the non-detections on September 24th,
making the reasonable assumption that the object was rapidly 
brightening before maximum, constrain the explosion epoch to 
September 25th (JD=2455099.5) $\pm 1$ day, which is one of the 
best constrained explosion epochs for a stripped-envelope SN 
(not associated with a GRB).

\begin{figure*}
   \includegraphics[width=17cm,height=9cm]{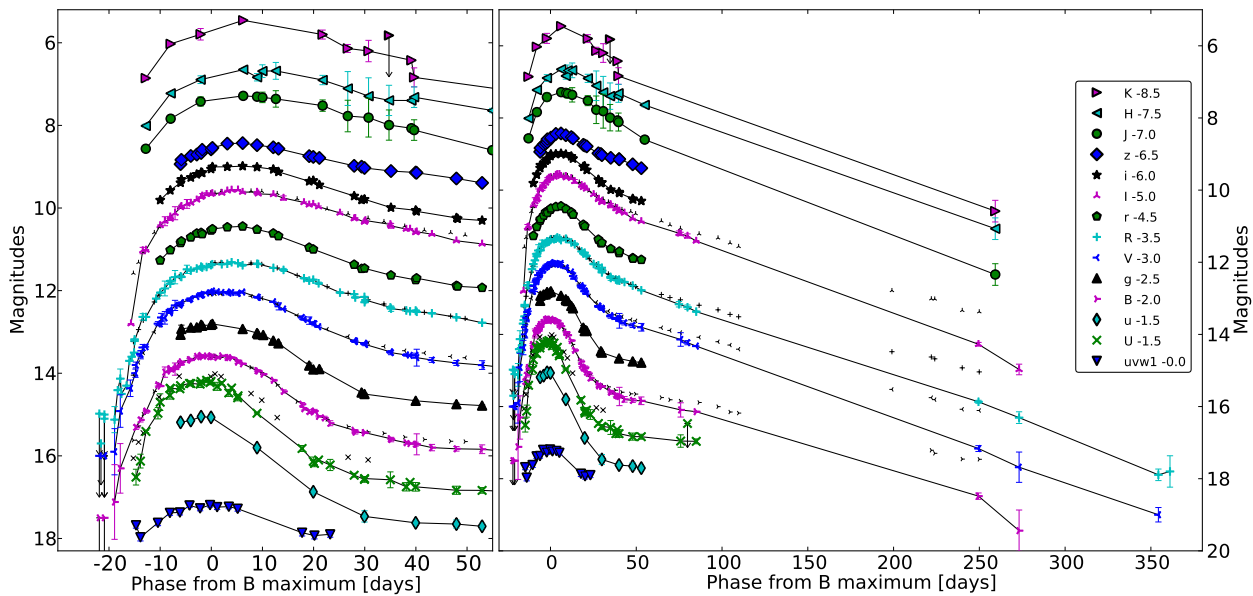}
  \caption{The light curves of SN 2009jf in the $uvw1-UuBgVrRiIzJHK$ bands. In the 
  left panel we show the data from explosion to two months after $B$ maximum 
  (JD=$2455119.4\pm1.0$, 2009 Oct. 14). In the right panel, we show the full light 
  curves in all bands. The data of \protect\cite{Sahu11} are overplotted with 
  smaller symbols for comparison.}
  \label{figLC}
\end{figure*}

All the collected photometric data are shown in Fig.~\ref{figLC}.
The data of \cite{Sahu11} are also overplotted for comparison.

As shown in Fig.~\ref{figLC} our photometry is consistent with that of
\cite{Sahu11} at early phases, but at later epochs their photometry
appears to systematically overestimate the SN flux as compared to our
template-subtraction photometry (see Appendix~\ref{ap0}).  
SN 2009jf is a clear case where proper template subtraction has to be 
used to avoid the contamination
from the bright background. There is also a shift between our $U$-band
photometry and that of \cite{Sahu11}, with our photometry being
fainter by $\sim0.2$ mag.  At late phases the differences are even
larger, again probably due to the overestimation of the SN magnitude
when using PSF-fitting techniques as opposed to template subtraction
photometry.

SN 2009jf reached its maximum luminosity
on 2009 October 14th at $15.56 \pm 0.02$ mag in the $B$-band.  It
peaked 2.4 days earlier in the $U$ band and 2.1, 4.6, and 5.3 days
later in the $V$, $R$, and $I$ bands respectively
(Table~\ref{tabparam}). The rise times to maximum were $\sim$17.5,
19.9, 22.0, 24.5 and 25.2 days respectively in the $UBVRI$ bands. This
is one of the slowest rise times ever observed for a classical SN
Ib/c.  Using a distance modulus of 32.65 $\pm$ 0.1 mag and a galactic
reddening of $E(B-V)$=0.112 mag \citep[][]{schlegel98} and internal
$E(B-V)$=0.05 mag (see Section~\ref{sec:host}), SN 2009jf reached a
maximum absolute magnitude in the $R$-band of $-18.24$ mag, brighter
than SNe 2008D \citep[$-17.26$ mag,][]{mazzali08} and 1999ex
\citep[$-17.78$ mag,][]{stritzinger02} and comparable with the massive
Type Ic core-collapse SN 2004aw \citep[$-18.22$
  mag,][]{taubenberger06}.  After maximum the light curves of SN
2009jf are not very different from those of other SNe Ib/c, although
with a slower than normally observed decline both soon after maximum
and after the inflection point at $\sim$30 days past maximum.  The
resulting light curve peak for SN 2009jf is broad, suggesting a
massive ejecta and/or a small expansion velocity which keeps the
ejecta optically thick for a long time.

The late-time luminosity decline rates were computed via weighted linear 
least-squares fits to the observations
and are reported in Table~\ref{tabparam} together with other key
parameters measured from the light curves.
The late-time slopes are in all bands steeper than those
expected if the energy source is $^{56}$Co~\to~$^{56}$Fe decay with
complete trapping of the $\gamma$-rays [0.98 mag
  (100~d)$^{-1}$] - as is normally observed in stripped-envelope 
  core-collapse supernovae.  

\begin{figure}
   \includegraphics[width=8.5cm,height=7cm]{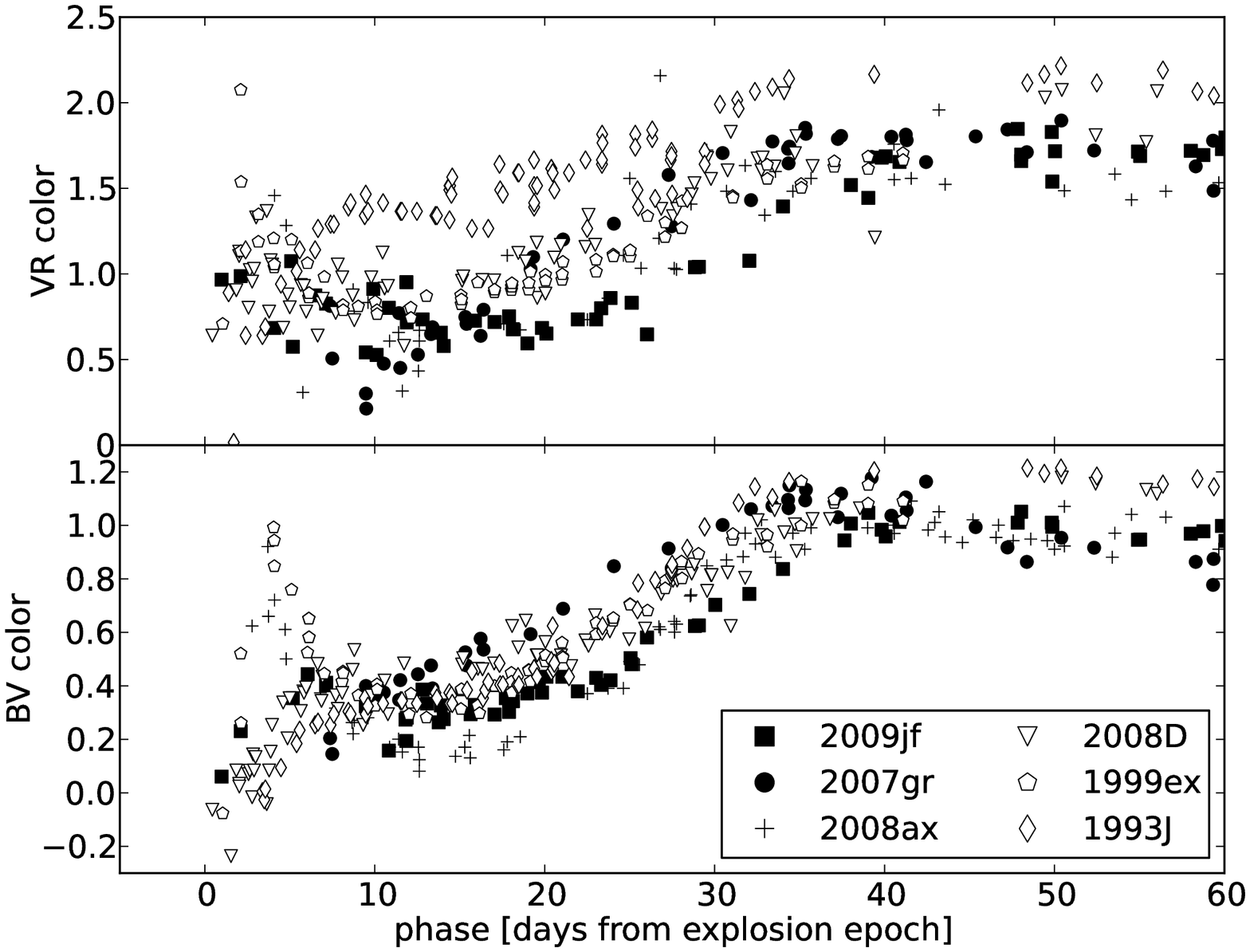}
  \caption{The $B-V$ and $V-R$ colour curves for a sample of stripped-envelope core-collapse supernovae.}
  \label{figcol}
\end{figure}

The colour evolution of SN 2009jf (see Fig.~\ref{figcol}) resembles
that of other stripped-envelope SNe. The SN becomes redder with time
as the ejecta expands and the temperature decreases. The $B-V$ colour
is 0.3 mag at +15 d from explosion and reaches a maximum $\sim$ +40 d
after explosion ($B-V$ $\sim$ 1 mag). In the first 2 weeks after
explosion some stripped-envelope SNe display a bluer $B-V$ colour
($B-V$ $\sim 0$ mag)
(e.g. SNe 1993J, 2008D). Other stripped-envelope SNe (e.g. SNe 1999ex,
2008ax), are blue immediately after explosion, show a $B-V$ colour
$\sim$ 0.5-1 mag at one week and then become blue again ($B-V$ $\sim$
0.3 mag) at two weeks after explosion. SN 2009jf shows a behaviour
similar to this second group of objects, but less extreme with $B-V$
$\sim 0$ mag in the first two photometric points, $B-V$ $\sim 0.5$ mag
at one week, and $B-V$ $\sim 0.3$ mag two weeks from explosion. The
blue colour of some SNe in the first days after explosion has been
interpreted as an evidence for shock break-out
\citep{stritzinger02,Chevalier10}.  With the caveat of the large
uncertainty in these first measurements, this may also be the case for
SN 2009jf.

\begin{table*}
 \centering
\begin{minipage}{180mm}
\caption{Main parameters for light curves of SN 2009jf. The $UBVRI$ magnitudes are in the Vega system, the $ugriz$ magnitudes in the AB system.}
\label{tabparam}
\begin{tabular}{cccccc}
\hline
Parameter   &  $U$ &  $B$ & $V$ & $R$ & $I$ \\
\hline
Date of max (JD $-$2,400,000)          & $55117. \pm 2.0$ &$55119.4\pm1.0$ &$55121.5\pm1.0$ &$55124.0\pm1.0$& $55124.7\pm1.0$\\
Apparent magnitude at max              & $15.70\pm 0.05$  &$15.56\pm0.02$        &$15.03 \pm 0.01$&$14.83\pm0.01$ & $14.59 \pm 0.01$\\
Absolute magnitude at max              & $-17.73\pm 0.26$ &$-17.75\pm0.22$      &$-18.12\pm 0.18$&$-18.24\pm0.15$& $-18.37\pm 0.14$\\
Late-time decline $\gamma$ (mag d$^{-1}$)& $-$    & $0.0088\pm0.0014$      &$0.0136\pm 0.0007$& $0.0161\pm0.0002$ &$0.0171\pm0.0011$\\
Phase range                            &  $-$     &     55-105                   &  63-105        &    60-105     & 58-105  \\
\hline
\end{tabular}
\begin{tabular}{cccccc}
\hline
Parameter   &  $u$ &  $g$ & $r$ & $i$ & $z$ \\
\hline
Date of max (JD $-$2,400,000)         & $55117.\pm 2.0$ & $55120.6\pm 1.0 $ & $55123.7\pm1.0$ & $55125.0 \pm 1.0$&$ 55125.5 \pm 1.0$  \\
Apparent magnitude at max             & $16.56\pm 0.04$ & $15.29\pm 0.01 $   & $14.96\pm0.02$  & $14.97 \pm 0.02$ &$ 14.95 \pm 0.03$   \\
Absolute magnitude at max             & $-16.88\pm0.26$ & $-17.97\pm 0.21 $  &  $-18.13\pm0.17$ & $-18.02\pm 0.14$ &$-17.94 \pm 0.12$   \\
Late-time decline $\gamma$ (mag d$^{-1}$)&$-$ & $0.015\pm0.002$  &  $0.020\pm0.003$&$0.021\pm0.002$&$0.016\pm0.001$    \\
Phase Range                                     &  $-$   &     49-73             &  55-73           &    55-73         & 48-73     \\
\hline
\end{tabular}
\end{minipage}
\end{table*}

\section{Photospheric spectra}
\label{parspe}

\begin{table*}
 \centering
  \begin{minipage}{140mm}
  \caption{Journal of spectroscopic observations}
  \label{tabseqspec}  
  \begin{tabular}{@{}cccccc@{}}
  \hline   
Date  &   JD  &  Phase
\footnote{Relative to $B$-band maximum light (JD = 2,455,119.4).} 
& Range & Resolution FWHM
\footnote{FWHM of night-sky emission lines.} 
&  Equipment \footnote{
 $~$A1.82 $=$ Asiago Ekar 1.82~m telescope; 
$~$ NOT $=$ Nordic Optical Telescope;
$~$  TNG $=$ Telescopio Nazionale Galileo; 
$~$ NTT $=$ ESO New Technology Telescope; 
$~$ CA $=$ Calar Alto 2.2m Telescope; 
$~$ WHT $=$ William Herschel Telescope;
$~$ VLT $=$ESO Very Large Telescope}
 \\
 &  2,400,000 & (days) & (\AA) & (\AA) & \\
 \hline
2009 Oct 02 & 55106.62 & $-13$ & 3200-10100 & 10  &  TNG+DOLORES+LRB/LRR \\
2009 Oct 04 & 55108.63 & $-11$ & 3600-9300  & 10  &  VLT+FORS2+300V/300I \\
2009 Oct 06 & 55111.40 & $-8$  & 3200-8750  & 10  &  CA+CAFOSC+b200 \\
2009 Oct 07 & 55112.37 & $-7$  & 6150-10100 & 10  &  CA+CAFOCS+r200 \\
2009 Oct 12 & 55116.57 & $-3$  & 3200-10100 & 10  &  TNG+DOLORES+LRB/LRR \\
2009 Oct 13 & 55118.46 & $-1$  & 3200-9200  & 15  &  NOT+ALFOSC+gr4 \\
2009 Oct 16 & 55121.48 & $+2$  & 3550-7750  & 23  &  A1.82+AFOSC+gr4    \\
2009 Oct 18 & 55123.42 & $+4$  & 3550-10000 & 23/35  &  A1.82+AFOSC+gr4/gr2 \\
2009 Oct 19 & 55124.61 & $+5$  & 3400-9650  & 14  &  NTT+EFOSC+gr11/16 \\
2009 Oct 27 & 55132.60 & $+13$ & 3600-9300  & 10  &  VLT+FORS2+300V/300I \\
2009 Nov 04 & 55140.39 & $+21$ & 3200-9200  & 15  &  NOT+ALFOSC+gr4 \\
2009 Nov 13 & 55149.39 & $+30$ & 3200-9200  & 17  &  NOT+ALFOSC+gr4 \\
2009 Nov 19 & 55155.35 & $+36$ & 3800-10000  & 12  &  CA+CAFOCS+g200 \\
2009 Dec 05 & 55171.34 & $+52$ & 3200-9200  & 15  &  NOT+ALFOSC+gr4 \\
2009 Dec 27 & 55193.35 & $+74$ & 3800-10000  & 12  &  CA+CAFOCS+g200 \\
2010 Jan 07 & 55149.39 & $+85$ & 3200-9200  & 15  &  NOT+ALFOSC+gr4 \\
2010 Jun 19 & 55366.88 & $+247$ & 3200-9200  & 18  &  NTT+EFOSC+gr11/16 \\
2010 Jul 08 & 55385.66 & $+266$ & 3200-9200  &  3   &  WHT+ISIS+R300B/R316R \\
2010 Oct 04 & 55473.57 & $+354$ & 3600-9200  & 25  &  NTT+EFOSC+gr13 \\
2010 Oct 11 & 55480.51 & $+361$ & 3600-9300  & 10  &  VLT+FORS2+300V/300I \\
\hline
2009 Oct 04 & 55109.51 & $-10$ & 8700-24700 & 18/35  &  TNG+NICS+IJ/HK \\
2009 Oct 21 & 55125.58 & $+6$  & 9400-24000 & 23/33  &  NTT+SOFI+GB/GR \\
2009 Nov 23 & 55158.58 & $+39$ & 8700-24700 & 18/36  &  TNG+NICS+IJ/HK \\
2009 Dec 05 & 55171.36 & $+52$ & 9300-24700 & 23/33  &  NTT+SOFI+GB/GR \\
\hline
\end{tabular}
\end{minipage}
\end{table*}

\begin{figure}
  \includegraphics[width=8.5cm,height=9.5cm]{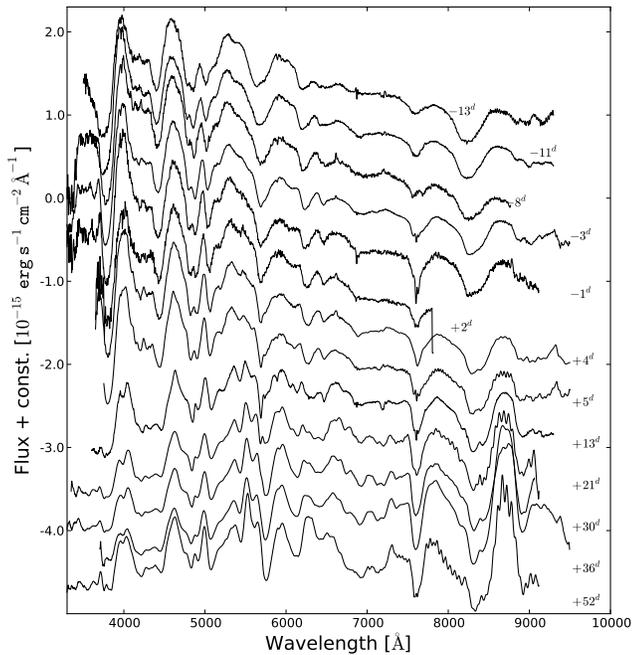}
  \caption{A subset of the collected photospheric spectra of SN 2009jf. The spectra are in the observer frame.}
   \label{fig:specevol}
\end{figure}

\begin{figure}
 \includegraphics[width=9cm,height=10cm]{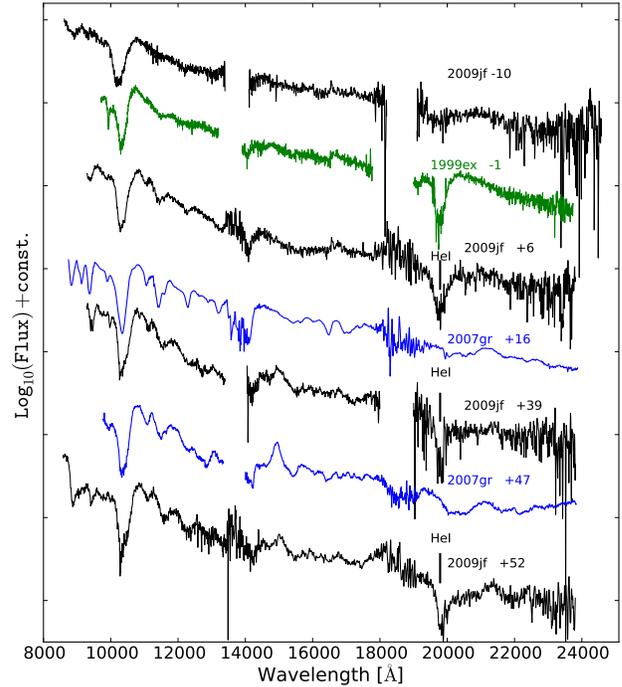}
  \caption{The infrared spectra of SN 2009jf are shown at the SN rest wavelength together 
  with those of SN 2007gr (Ic) and SN 1999ex (Ib). \HeI{} \L{}2.058 $\micron$ is marked 
  in the spectra of SN 2009jf.}
   \label{fig:speinfrared}
\end{figure}

A subset of the optical photospheric spectra of SN 2009jf is shown in
Fig.~\ref{fig:specevol}, the instrumental configurations used at each
epoch are listed in Table~\ref{tabseqspec}, while details of data
reduction are in Appendix~\ref{ap0}. The four NIR spectra,
are shown in Fig.~\ref{fig:speinfrared} together with spectra of other
stripped-envelope SNe. The first spectrum of SN 2009jf was observed 13
days before maximum (one week after explosion).  At early phases, SN
2009jf shows a blue continuum with the typical features of a
stripped-envelope SN already visible, namely \FeII{}
\LL{}4924,5018,5169, \CaII{} \LL{}8498,8542,8662, \CaII{}
\LL{}3934,3968 and \OI{} \L{}7770.  The lines are partially blended,
but not as much as for broad-lined Ic (BLIc) SNe or the energetic Ib
SN 2008D \citep{mazzali08}.

With time, the temperature decreases, the spectra of SN 2009jf become
redder and lines which form deeper in the ejecta (at lower velocity)
become more prominent.  The identification of the features at $\sim$
5800 \AA{} and $\sim$ 6200 \AA{} is more complicated as these regions
are usually densely populated with lines from several different
elements such as Na, He, H, Si, C and Ne.
 
The possible presence of H and He is very important  to understand the 
progenitor star evolution. In more massive progenitor (M $>$ 25-30 \msun{}) 
the ratio between He (or H) and the other elements of the ejecta should 
be smaller than in less massive progenitors (11 \msun{} $<$ M $<$ 25 \msun{} ).
This will be discussed in the following section.

\subsection{He identification}
\label{sec:He}
The He lines, with the exception of the \HeI{} line at 2.058
$\micron$, are in spectral regions which are densely populated by
other features.  The identification of He lines is further complicated
by the fact that they often form in non-LTE conditions, giving
uncertain line strength ratios \citep{lucy91}.

\begin{figure}
 \resizebox{\hsize}{!}{\includegraphics{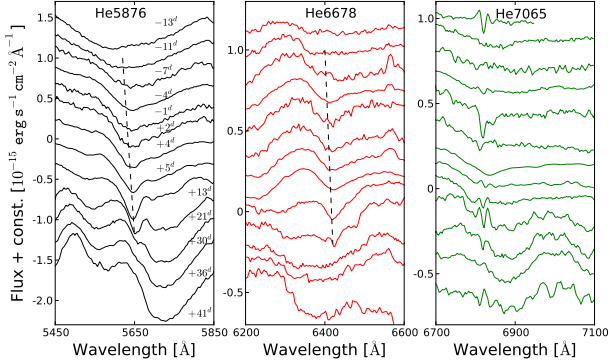}}
  \caption{The spectral regions where the He lines (at \L 5876, \L 6678 and \L 7065) should be visible (spectra at rest frame).  
  The absorption at $\sim$ 5700 \AA ~ visible at late phases is unlikely to be He, as suggested by \protect\cite{Sahu11} 
  since all other He lines at that velocity are missing both in the optical and in the near infrared.}
   \label{fig:Helium}
\end{figure}

For some recent detailed studies of SNe Ib/c the detection or
non-detection of the \HeI{} line at 2.058 $\micron$ was the clearest
way to distinguish SNe Ib from SNe Ic
\citep{hamuy02a,valenti08b,stritzinger09,modjaz09}.  For SN 2009jf,
\HeI{} at 2.058 $\micron$ is clearly detected in our NIR spectra at 6, 39 
and 52 days after maximum and marginally detected also at 
10 days before $B$-band maximum (see Fig.~\ref{fig:speinfrared}). 
Thus He is definitely present in SN 2009jf, confirming the SN classification 
as a SN Ib. However, while for most SNe Ib the He lines increase in intensity 
from the explosion until two weeks after maximum, in SN 2009jf the He lines 
are weaker than in other SNe Ib, with the lines at \L{}6678 and \L{}7065 almost
disappearing two weeks after $B$-band maximum.  In
Fig.~\ref{fig:Helium}, we show the spectral regions where He lines
should be visible.  The \HeI{} \L{}5876 and \L{}6678 are clearly seen
in Fig.~\ref{fig:Helium} (minimum of the narrow P-Cygni absorption is
indicated with a dashed line). On the red side of the He line at
\L{}5876, a broader absorption becomes more intense over time. This
feature was identified by \cite{Sahu11} as He at much lower velocity
than at early phases. However, no sign of such absorption is visible
for the He lines at \L{}6678, \L{}7065 and 2.058 $\micron$. In order to better
constrain the presence of He, we used the {\sc synow}\footnote{
  http://www.nhn.ou.edu/$\sim$parrent/synow.html} spectrum synthesis
code to model two spectra of SN 2009jf.

\subsubsection{-11 d from $B$-band maximum}
\begin{figure}
  \includegraphics[width=9cm,height=10cm]{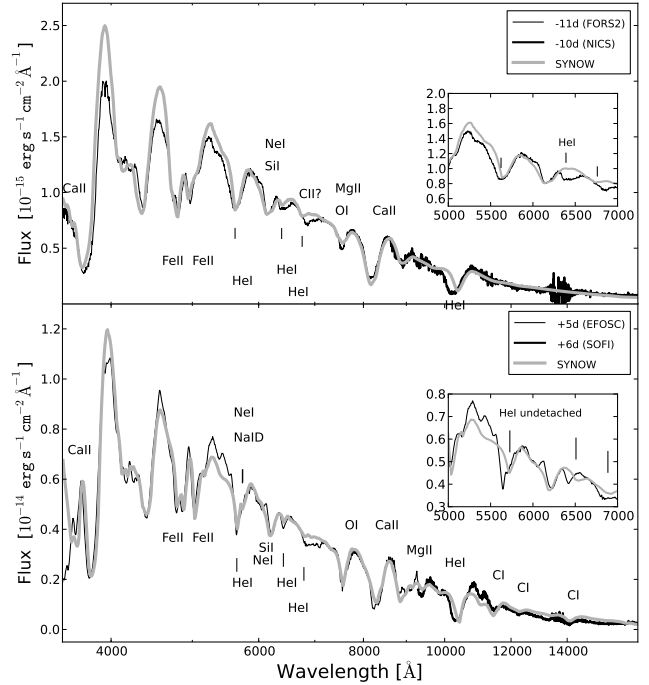}
  \caption{{\sc synow} fit of the spectra of SN 2009jf at -11 days and +6 days from $B$-band  maximum light. 
  Before maximum the He lines are present in the {\sc synow} fit, while in the inset, using Na instead of He, 
  the line at $\sim$6500 \AA{}, which is likely \HeI{} \L{}6678 is not reproduced. 
  After maximum  most of the line at 5700 \AA{} is reproduced by \NaID. Only undetached lines of He are identified.
 The inset shows a {\sc synow} spectrum with undetached He. The spectra are at rest frame.}
   \label{fig:synow}
\end{figure}

We have modelled the merged optical (VLT+FORS2) and near infrared
(TNG+NICS) spectra taken $-11$ and $-10$ days before $B$-band maximum
respectively.  The spectrum was reproduced including lines from \FeII{},
\OI{}, \CaII{}, \MgII{}, \ScII{}, \TiII{}, \SiII{}, \NeI{} and \HeI{}
(upper panel Fig.~\ref{fig:synow}).  We used a photospheric velocity
of 13500 \kms{} and line optical depths that vary as ${e^{-v/v_{e}}}$
with $v_{e}$ = 2000 \kms{}. The features at 6100 \AA{} were well
reproduced by a combination of \SiII{} and \NeI{}, while \HeI{}
(undetached\footnote{lines that form at the photospheric velocity}) is
able to reproduce both the feature at $\sim$5600 \AA{} and the one at
$\sim$6500 \AA{}.  The presence of \CII{} can not be ruled out, but
while \CII{} \L{}7235 would improve the fit of the absorption at
$\sim$ 7000 \AA{}, the \CII{} \L{}6580 line would deteriorate the fit
at $\sim$ 6400 \AA{}.  At this epoch, the \HeI{} at 2.058 $\micron$ is
not visible in the NIR spectrum (see Fig.~\ref{fig:speinfrared}). For
this reason, we also considered the possibility that the absorption at
$\sim$ 5600 \AA{} is \NaID{} instead of \HeI{}.  However, with Na,
instead of He, the fit will also fail to reproduce the features at
$\sim$6500 \AA{}, which is likely \HeI{} (\L{}6678) (see inset in
upper panel in Fig.~\ref{fig:synow}).  
Helium is also visible  at 2.058 $\micron$ (see Fig. \ref{fig:speinfrared}), 
though not very strong. We note that early NIR spectra
are also available for another SN Ib (SN 2008D), in which the \HeI{}
at 2.058 $\micron$ is also marginally visible at high velocity 
also at early phases \citep{modjaz09}.
We hence consider the He identification quite robust. 

\subsubsection{+5 d from $B$-band maximum}

The NTT+EFOSC (+5 d) and NTT+SOFI (+6 d) spectra and {\sc synow} fit
are shown in the lower panel of Fig.~\ref{fig:synow}.  We used the
same ions as for the pre-maximum spectrum, plus \NaI{} and \CI{} at a
velocity of 8200 \kms. \CI{} is needed to fit some of the features
between 10000 and 16000 \AA{}, while \NaID{} is needed to reproduce
part of the spectrum at $\sim$ 5600 \AA{}. \HeI{} is mainly
responsible for the blue narrow component of this feature with a
detached velocity of 12000 \kms{}. The detached He is also responsible
for the line at $\sim$ 6500 \AA{}.  We investigated the possible
presence of undetached \HeI{}, but in this case the He velocity is too
low to reproduce correctly the absorbtion at $\sim$ 6500 \AA{}
(see lower inset panel in Fig.~\ref{fig:synow}).  Also, as we will see
in the next section, the photospheric velocity at which the \HeI{}
feature at 2.058 $\micron$ forms (measured from the minima of the
P-Cygni absorption) are consistent with detached He. The photospheric
velocity of \HeI{} at 2.058 $\micron$ remains constant from +6 to +52
days after $B$-band maximum.

\begin{figure}
  \includegraphics[width=9cm,height=9cm]{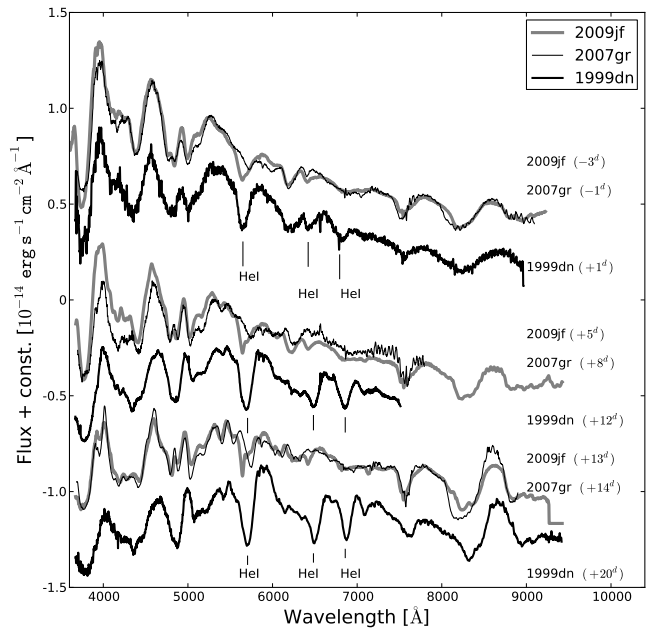}
  \caption{Three spectra of SN 2009jf are compared with those of SNe 2007gr and 1999dn. 
  At all epochs the spectra of SN 2009jf are almost identical to those of SN 2007gr, 
  with the sole exception of the He lines. SN 1999dn shows strong He 
  absorption features, in contrast to both SNe 2009jf and 2007gr. All spectra are at rest frame.}
   \label{fig:comparison}
\end{figure}

Summarising the optical and infrared spectral analysis, we can
conclude that the layer where the He lines form is located at a
velocity between 12000 and 16000 \kms{}. It appears undetached at early
phases, but becomes detached at $1-2$ weeks after maximum.  This is
not surprising, as similar behaviour has already been observed in
other SNe Ib \citep{branch02}. Regarding the presence of H, while a
small amount cannot be completely ruled out, in both spectra there is
no clear improvement in the fit when H is included. Moreover, there is
no evidence (as observed for some SNe Ib) of \Hb{} features in the
early spectrum. Hence SN 2009jf is likely a H-poor stripped-envelope
SN, with a relatively small amount of He (mainly detached).  This is
confirmed by comparing the spectra of SN 2009jf with other SNe Ib/c
(see Fig.~\ref{fig:comparison}).  Across a range of phases, SN 2009jf
is almost identical to the He-poor SN 2007gr
\citep{valenti08a,hunter09}, with differences only in the region of
the most prominent He features.  For comparison, some spectra of the
Ib SN 1999dn \citep{benetti11} are shown with more prominent He
features.  Apart from the Ib classification, we could also regard SN
2009jf as a SN Ic that exploded with a small amount of He left in a
high velocity shell at 12000-16000 \kms{}.

\subsection{Photospheric velocity}
\label{parvelosity}

\begin{figure}
  \includegraphics[width=8.5cm,height=11cm]{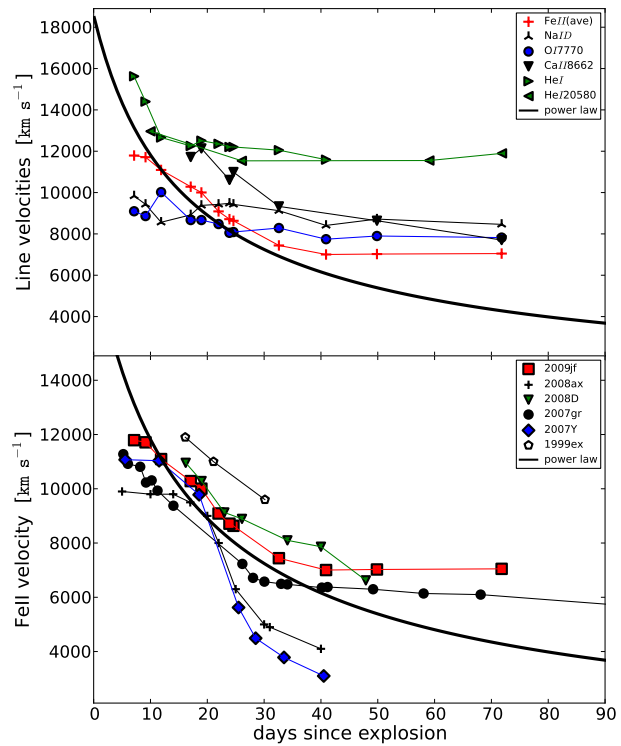}
  \caption{Upper panel: Line velocities of SN 2009jf. The \NaID{} and \HeI{}  minima 
  have been measured by simultaneously fitting two gaussians to the absorption feature at $\sim$ 5600 \AA{}. 
  Lower panel: The photospheric velocity evolution of SN 2009jf in comparison with those of a set of stripped-envelope SNe. All velocities were measured in {\sc iraf} with a Gaussian fit to the minimum of the line. In the both panel the solid curve show the power-law fit to the photospheric velocities of a sample of SNe Ib from \protect\cite{branch02}.}
   \label{fig:velocevol}
\end{figure}

In the previous section we claimed the presence of a detached layer of
He at $\sim$ 12000 \kms{}. This is also evident in
Fig.~\ref{fig:velocevol} (upper panel), where the velocity evolution
for different ions is plotted. The isolated \HeI{} line at $\sim$ 2
$\micron$ confirms the presence of detached He, while there is no sign
in the infrared spectra of He at low velocity.

Different lines form in different regions of the ejecta, in part due
to the differing physical conditions (temperature and density), 
but also due to the stratification of elements
within the ejecta. The \CaII{} lines form in the outer parts of the
photosphere (as can be seen from their higher velocities), while O and
Fe lines form in the inner part. The \FeII{} lines usually provide a
good estimate of the photospheric velocity. Comparing with other
stripped-envelope SNe, SN 2009jf has quite a high photospheric
velocity, slightly lower than SNe 1999ex and 2008D but higher than SNe
2007Y and 2008ax.  In particular at late phases (70 days after
explosion), SN 2009jf still has an optically thick photosphere at 7000
\kms{}. The evolution of the photospheric velocity of SN 2009jf
resembles that of SN 2007gr, but with velocities which are $\sim$1000
\kms{} greater at almost all epochs. The similarity in the spectra
(see previous section) and in the photospheric velocity
evolution suggests that SN 2009jf had a progenitor similar to that of
SN 2007gr, albeit slightly more massive and with more He left at the
time of the explosion.
    
\section{Nebular spectra}
\label{spenebular}

\begin{figure}
 \resizebox{\hsize}{!}{\includegraphics[width=9cm,height=10cm]{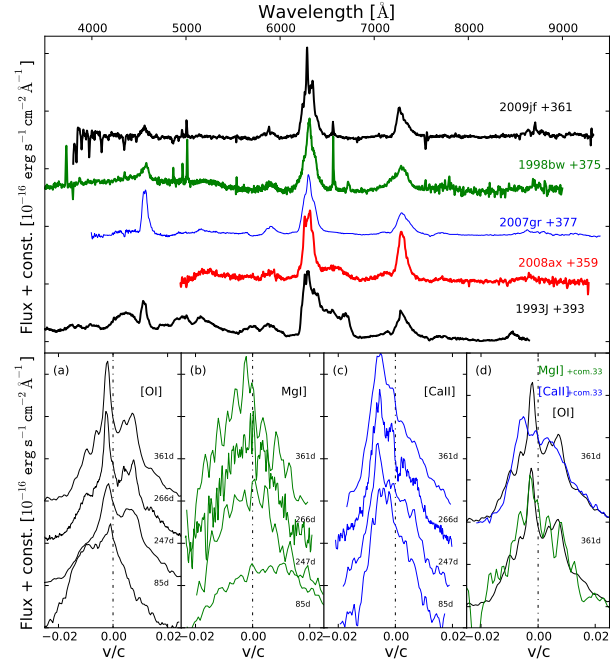}}
  \caption{\textbf{Upper panel}: Nebular spectra (at rest frame) of a sample of stripped-envelope 
  SNe at 1 year from $B$-band maximum. \textbf{Bottom panel}. Evolution of the main nebular 
  features visible in the nebular spectra of SN 2009jf: (a) [\OI], (b) \MgI] and (c) [\CaII]. 
  In panel (d) we show at 361 days after $B$-band maximum,  the comparison 
  between the [\OI] profile and profiles of the [\CaII] and \MgI]  modified
  with a second component to reproduce the [\OI] doublet (see text).}
   \label{fig:nebularspectra}
\end{figure}

After SN 2009jf became visible again in June 2010, we obtained spectra
at 247, 266, 354 and 361 days after $B$-band maximum (see
Fig.~\ref{fig:nebularspectra}).  At these late epochs the SN ejecta
are optically thin, allowing us to observe the innermost parts of the
ejecta where lines are mainly in emission.  Several studies have been
performed on nebular spectra of stripped-envelope SNe to investigate
the geometry of the explosion
\citep{mazzali05,maeda08,modjaz08b,taubenberger09} as the nebular
lines approximately represent a one-dimensional line-of-sight
projection of the three-dimensional distribution of elements.

The most prominent emission lines are the [\OI] \LL{}6300,6364
doublet, \MgI] \L{}4571 and [\CaII] \LL{}7291,7323. Several features
  of \FeII\/ are also visible (but highly blended) from 4500 to 5500
  \AA\/.  These are prominent only in the brightest stripped-envelope
  SNe \citep[e.g. in SN 1998bw,][]{mazzali01c}. The [\OI] doublet is
  best suited to probe the explosion geometry, as the [\CaII] feature
  can be partially contaminated by [\OII] \LL{}7320,7330, and the
  \MgI] line is less intense and contaminated by Fe lines (at $\sim$
    100 days after explosion).  Oxygen has also the advantage that it
    is the most abundant element in strongly stripped core-collapse
    SNe.  A drawback is that it is a doublet with a line ratio
    sensitive to the temperature and density where the lines form.

It has been predicted and observed that the ratio of the oxygen lines
\L{}6300/\L{}6364 increases with time from a ratio of 1:1 to a ratio
of 3:1 in H-rich SNe \citep{chugai92,leibundgut91,spyromiglio91}.  For
stripped-envelope SNe, \cite{taubenberger09} suggested that the
conditions during the nebular phase in the oxygen layer always 
give a ratio of 3:1.
\cite{Milisavljevic10}, on the other hand, suggested that the [\OI]
line ratio may be close to 1:1 in several of these SNe, in order to
explain the fact that SNe Ib show a double peak for the O feature more
often than SNe Ic (without invoking a highly asymmetric
explosion). Recently, \cite{hunter09} and \cite{taubenberger11}
demonstrated that for SN 2007gr and SN 2008ax magnesium and oxygen had
a similar distribution within the ejecta if the ratio of the oxygen
lines at \L{}6300/\L{}6364 was fixed at 3:1.  This similar
distribution is expected from nucleosynthesis models
\citep{maeda06a}. Furthermore, \cite{maurer10} suggested that the
apparent double peak at the position of the oxygen doublet is due to a
high-velocity \Ha{} absorption which, when superimposed on the oxygen
lines, gives the appearance of a double peak.  All SNe in the sample
analysed by \cite{maurer10} are SNe IIb.

SN 2009jf also shows an oxygen feature with a complex structure that
resembles a double peak profile. But since SN 2009jf does not show 
H features either at early or nebular phases, we consider unlikely that
the [\OI] profile of SN 2009jf is due to contamination from
high-velocity H.  This is also supported by the comparisons with \MgI]
at \L{}4571 and [\CaII] at \LL{}7291,7323 on day +361 : if we take into account
the doublet nature of the [\OI] and artificially add a second
component (redshifted and scaled to 1/3 of the flux) to the \MgI] and
[\CaII] lines \footnote{ As the [\CaII] feature is a close doublet,
it has been considered here as a single line.}, 
then the profiles of [\OI], modified-\MgI] and modified-[\CaII]
  all have similar profiles as shown in the bottom panel ($d$) of
  Fig.~\ref{fig:nebularspectra}.  The similar profiles suggest that
  mixing is important in the progenitor of SN 2009jf since oxygen,
  magnesium and calcium have similar distributions within the
  ejecta, and that the oxygen profile is not due to \Ha{}
  contamination.

\begin{figure}
 \resizebox{\hsize}{!}{\includegraphics[width=8cm,height=14cm]{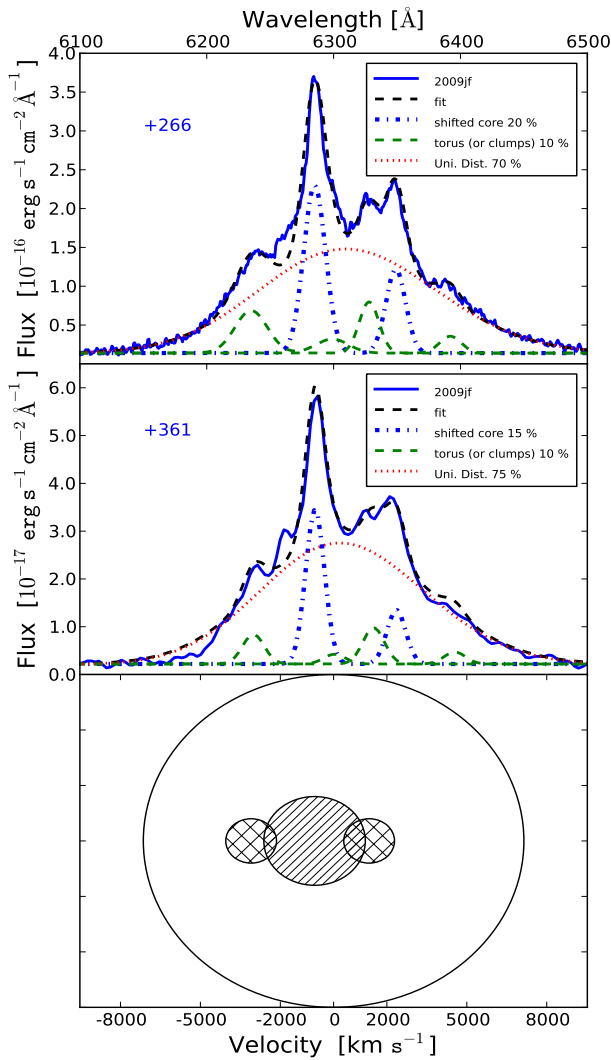}}
  \caption{Oxygen line profile of SN 2009jf at +266 and +361 days from B 
maximum (upper and middle panel). If we assume that the oxygen line flux to trace 
  the oxygen mass, SN 2009jf could be an asymmetric explosion, with an 
  off-axis core ($\sim$ 20$\%$ of the oxygen flux/mass) surrounded by clumps.
 The remaining oxygen  ($\sim$ 70$\%$) is distributed uniformly throughout the ejecta. 
In the lower panel a schematic reconstruction of the geometry is shown.}
   \label{fig:nebularspectra2}
\end{figure}

If we assume that the mass of an element at a particular velocity is
simply proportional to the emitted flux at that same velocity, the O
profile (as those of Mg and Ca) is consistent with an
asymmetric explosion, with a large part of the material ejected away
from the observer.  \cite{Sahu11} also discussed the nature of the
oxygen profile, and suggested large-scale clumping or a unipolar jet
as the explanation.  Here we propose an alternative geometry, which
reproduces the oxygen profile with four different components (see
upper panel of Fig.~\ref{fig:nebularspectra2}) coming from different
parts of the ejecta. Taking into account that the [\OI] line is a
doublet (two lines separated by with a 64 \AA{} and with a ratio $\sim$ 3:1), there
are two clear narrow lines at 6285 \AA{} and 6349 \AA{} in the oxygen
profile that may be interpreted as an off-centre dense core that is
blueshifted by $\sim$ 700 \kms{}. The ratio of these two lines seems
slightly lower than 3:1 suggesting a high density in this blob.
Blueward and redward by $\sim$ 50 \AA{} with respect to the off-center
dense core, we identify further blobs\footnote{the two components of
  the blobs have ratios of $\sim$ 3:1} of oxygen-rich material that
may be interpreted as clumps. 
In this scenario the rest of the oxygen (the broader component of the O line) 
is uniformly distributed in the ejecta\footnote{This is reproduced with 
two Gaussians (ratio $\sim$ 3:1) close to zero velocity.}
A schematic reconstruction of the geometry is shown in 
Fig.~\ref{fig:nebularspectra2} (lower panel).
Within our simplifying assumption of mass proportional to flux, the
dense off-center core would contain $\sim$ 20$\%$ of the oxygen mass,
the clumps $\sim$ 10$\%$, while the rest of the oxygen mass
would be distributed uniformly. We performed this analysis on the
spectra at both +266 days and +361 days after $B$-band maximum,
obtaining similar results, as there is no significant evolution in the
nebular spectra over this time period.

A caveat on the above discussion should be added. While the
 assumption that the oxygen mass distribution is proportional to the
 oxygen lines flux is reasonable, the random directions of material ejection
 in asymmetric explosions should produce stripped-envelope SNe with
 observables (e.g. line profiles) isotropically distributed.
 Inspecting the sample of nebular spectra of \cite{taubenberger09}
 (and including some recent discoveries), a number of SNe have been found
 showing blue-shifted, off-center line cores, while SNe with red-shifted
 off-center line cores have not been detected so far.
 This investigation is complicated by the fact that the oxygen profile may
 show blue-shifted line peaks also at early phases ($<$ 200 days, when the
 ejecta are not fully transparent) or at very late phases due to dust
 formation.
 However, the amount of blue-shift of the oxygen line peaks in the above
 mentioned spectra do not show any time evolution, leading us to rule out
 the dust formation or still opaque ejecta as likely explanations.
 Nevertheless the lack of detections of SN spectra with red-shifted oxygen
 profiles is puzzling and an obvious explanation cannot be provided.

\section{Host galaxy properties}
\label{sec:host}
\begin{figure}
\center
   \includegraphics[width=6cm,height=5cm]{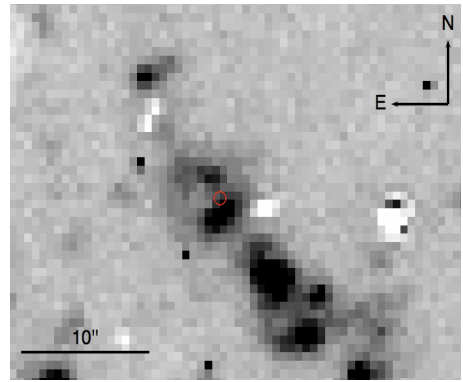}
  \caption{Continuum-subtracted H$\alpha$ image of NGC7479 with the SN position indicated with a circle. 
  The radius of the circle is 10 times the uncertainty in transformation used to determine the SN position.}
  \label{figha}
\end{figure}

The host galaxy of SN 2009jf, NGC 7479 is a face-on spiral galaxy in
the Pegasus constellation. It is relatively nearby ($\mu=32.65$
mag)\footnote{Using a radial velocity corrected for in-fall onto Virgo
  of 2443 \kms{} and a Hubble constant of 72 \kms{} Mpc$^{-1}$.}, and
quite asymmetric, with strong star formation along most of the
luminous western arm \citep{laine98}. SN 2009jf occurred in this arm,
at the location of an extended star forming region (see
Fig.~\ref{figha}). NGC 7479 also displays some intriguing properties
at radio wavelengths, with a radio continuum in the reverse direction
to the optical arms \citep{laine05}, which has been suggested to be
the result of a minor merger.

\begin{figure}
\center
   \includegraphics[width=6cm,height=6cm]{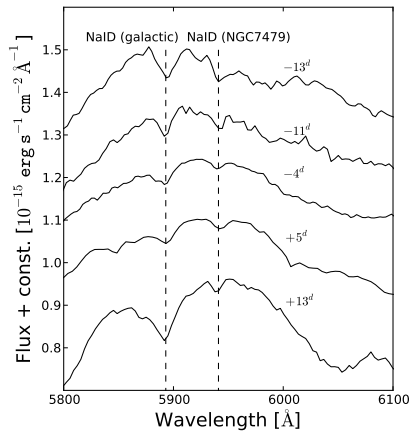}
  \caption{Low resolution spectra of SN 2009jf in the region of the Galactic 
  and host-galaxy Na~{\sc i}~D absorptions}
  \label{fignaid}
\end{figure}

The early spectra and the colour evolution of SN 2009jf are quite
blue, suggesting that the light from SN 2009jf is absorbed and
reddened by a relatively small amount of interstellar dust both in the
Milky Way and in the host galaxy NGC 7479. While the Milky Way
component is easily removed using maps of the Galactic dust
distribution and a standard extinction law \citep[$E(B-V)$ = 0.112
  mag, ][]{schlegel98}, evaluating the extinction in NGC 7479 is more
difficult. In principle, assuming an average dust-to-gas ratio, the
reddening can be estimated by measuring the gas column density of
interstellar lines from Na~{\sc i}~D absorption. It is common practice
to derive the host galaxy extinction from its relation with the
equivalent width (EW) of the Na~{\sc i}~D line.  Mostly using a sample
of SNe Ia for which the reddening has been calculated using the Lira
relation \citep{phillips99}, \cite{turatto03} found that SNe appear to
split  on two slopes that give quite different EW(NaID) versus
reddening relations. For SN 2009jf, we used the early spectra with
high signal-to-noise ratio in order to separate the contribution to
the Na~{\sc i}~D absorption of the Milky Way from that of host system.
Two small Na~{\sc i}~D absorptions are visible in several spectra at
5893 \AA{} (Galactic) and at 5941 \AA{} (host system) (see
Fig.~\ref{fignaid}). The EW measured for each absorption slightly
changes among the spectra (0.4-0.9 \AA{} for the Galactic absorption
and 0.2-0.4 \AA{} for the one in the host system). This is probably
due to the different signal-to-noise ratios in the spectra, the
spectral resolution and/or contamination from other lines.  On the
other hand, the ratio of the two absorptions is almost constant, with
the Galactic one being twice as strong as that in the host system.
Assuming a similar dust-to-gas ratio in NGC 7479 and in the Milky Way and
the Galactic reddening in the direction of SN 2009jf reported by
\cite{schlegel98}, we obtain an $E(B-V)_{7479} \sim$ 0.05 mag. 
Due to the uncertainty of the method and the fact that the EW measurement
is uncertain, we adopt a host-galaxy reddening of $E(B-V)_{7479}$= 0.05 $\pm$
0.05 mag.

\begin{table}
\caption{Main parameters for SN 2009jf and its host galaxy}
\label{tabsummary}
\begin{tabular}{ll}
\hline
Parent galaxy           &  NGC 7479             \\
Galaxy type             &  SBbc$^a$             \\
RA (2000)               & $23^h 04^m 52^s.98$   \\
Dec (2000)              & $+12\degr 19' 59.5''$ \\
Recession velocity      & 2443 [\kms] $^a$      \\
Distance modulus ($H_0 = 72 $) & $32.65 \pm 0.10$ mag \\
$E(B-V)_{7479}$         & 0.0 - 0.1  mag \\
$E(B-V)_{MW}$          & 0.112 mag$^b$ \\
Offset from nucleus     & $53''.8$ W , $36".5$ N \\
Explosion epoch (JD)   & $2455099.5\pm 1.0$ (Sep 25, 2009) \\
\hline
\end{tabular}\\
$^a$LEDA, velocity corrected for Local Group infall onto the Virgo cluster.\\ 
$^b$\protect\cite{schlegel98}.
\end{table}

\subsection{Metallicity}

NGC 7479 has a bright $B$-band absolute magnitude of
$M_{B}$ = $-21.64$\footnote{HyperLEDA; http://leda.univ-lyon1.fr/}.  Using the
calibration of \cite{Boissier09}, suggests a super-solar oxygen
abundance of 12 + log(O/H) = 9.09 dex at the characteristic radius of
0.4 R$_{25}$\footnote{The R25 radius is the radius of the 25 mag
  arcsec$^{−2}$ B-band isophote.}.  A direct estimate for the host
galaxy metallicity can be obtained from the line fluxes of nebular
emission lines in the vicinity of the SN.  Actually, many different
diagnostics for metallicity can be found in the literature \citep[for
  example ][]{pettini04,kewley02,mcgaugh91,Pilyugin05} using various
emission-line ratios and calibrations.  Different metallicity
calibrations also give systematically different results
\citep{smartt09a,ellison05,modjaz08a}.  In this paper, we follow
\citeauthor{modjaz08a} and use a range of methods to determine the
metallicity.

\begin{figure}
  \includegraphics[width=9cm,height=6cm]{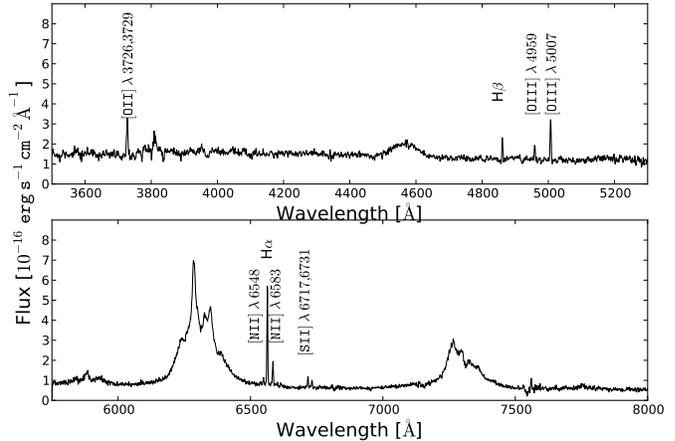}
  \caption{Host-galaxy emission lines in the late-time spectrum of SN 2009jf 
  at +266 days from $B$-band maximum.}
   \label{fig:fig_neb_spec}
\end{figure}

The strengths of the host-galaxy emission lines were measured in the
deep spectrum of SN 2009jf obtained on 2010 July 7th with the WHT and
ISIS (see Fig.~\ref{fig:fig_neb_spec}).  The measured fluxes are
listed in Table~\ref{tab_neb_lines}.

\begin{table}
\center
\caption{Nebular emission lines seen in the spectrum of 2010 July 7th. 
The spectrum has been corrected for host and Galactic extinction, and 
the fluxes of all emission lines of interest were measured subtracting the continuum
and fitting a Gaussian to the line using {\sc iraf}.}
\label{tab_neb_lines}
\begin{tabular}{llr}
\hline
Species 		& Wavelength	& Flux ($10^{-17}$)\\
				& (\AA)		& (\ene $cm^{-2}$)  \\
\hline		
[\OII{}]		    & 3726,29	& 	183	\\
H$\beta$		& 4861		&	58	\\
$[$\OIII{}$]$  	        & 4959		&	45	\\
 $[$\OIII{}$]$         	& 5007		&	125	\\
$[$\NII{}$]$ 		        & 6548		&	20	\\
H$\alpha$		& 6563		& 	273	\\
$[$\NII$]$		& 6583		&	68	\\
$[$\SII$]$		& 6717		&	31	\\
$[$\SII$]$		& 6731		&	25	\\
\hline
\end{tabular}\\
\end{table}

The N2 index calibration of \cite{pettini04} gives 12+log(O/H) = 8.56
dex, while the O3N2 index by the same authors gives a value of 8.43
dex. \cite{Pilyugin11} use the NS calibration to give relations for
the oxygen abundance 12+log(O/H), and the nitrogen abundance
12+log(N/H). With the measured fluxes near the site of the SN we
derive 8.36 dex for the former and 7.49 dex for the
latter. \cite{Kobulnicky04} give an approximation for the average of
the KD02 and M91 relations, which gives 12+log(O/H) = 8.67 dex.

The mean of these four values for the metallicity is 8.51 dex, which
is between the solar value \citep[8.72 dex, ][]{allende2001} and that
of the Large Magellanic Cloud \citep[8.35 dex, ][]{hunter07} 
which is considerably smaller than the value obtained with the $B$-band absolute
magnitude of NGC 7479.

\subsection{The progenitor of SN 2009jf}
\label{parprogenitor}

To search for possible evidence of the progenitor, the post-explosion
NaCo $K_\mathrm{S}$-band image from October 2009 was aligned with the stacked
pre-explosion WFPC2 $F814W$ image. 12 sources common to both images
were identified and their positions were measured accurately with the
{\sc iraf phot} task. Using the resulting list of matched pixel
coordinates, we derived a geometrical transformation\footnote{As a
  small number of common sources were used for the alignment, we
  restricted the transformation to rotation, scaling and translation
  only.}  between the two images, using {\sc iraf geomap}.  The rms
error in the transformation was found to be 0.573 WF pixels, or 57
mas.

\begin{figure*}
 \resizebox{\hsize}{!}{\includegraphics{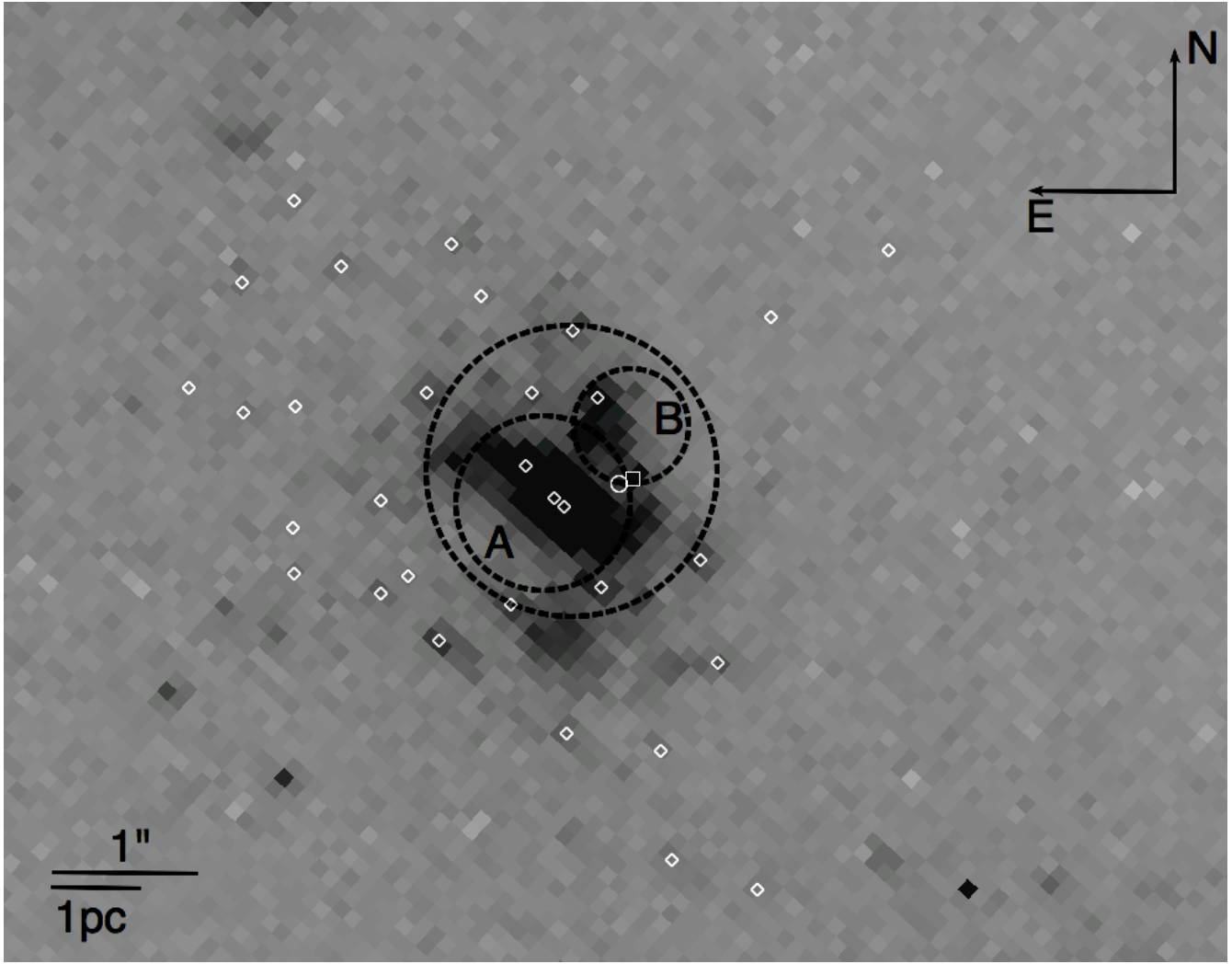}}
  \caption{The $HST$+WFPC2 $F569W$ image of the location of SN 2009jf. 
Scale and orientation are indicated. The position of the SN, 
as determined from the post-explosion NaCo image is located with a white 
circle. The radius of the circle corresponds to the 57 mas uncertainty in the 
SN coordinates and the geometric transformation. All sources detected by 
{\sc hstphot} within a distance of 31 pixels of the SN location (corresponding 
to a 500 pc radius at the distance of NGC 7479) at the 3$\sigma$ level in both 
filters are indicated with white diamonds. The source located close to the SN 
in the $F569W$ filter is indicated with a white square. The regions used for 
aperture photometry of the complexes as discussed in the text are indicated 
with dashed circles, and labelled accordingly. }
\label{fig_F569}
\end{figure*}

The position of the SN was measured in the NaCo image using the three
different centering algorithms in {\sc iraf phot}, all of which agreed
to within $\sim$ 5 mas.  The mean of the three positions was taken as
the SN position, this was then transformed to the pixel coordinates of
the WFPC2 $F814W$-filter image\footnote{The derived pixel position of
  the SN in the coordinates of u2z00103t.c0f.fits are
  471.55,134.53.}. The SN is in a crowded region, dominated by two
bright complexes (A and B) one of which appears elongated (see
Fig.~\ref{fig_F569}).  We also aligned the $F569W$ image using 
the same procedure\footnote{The $F569W$ image was found
  to be offset from the $F814W$ image by ($-0.05,-0.43$) WF pixels in
  x and y respectively, this offset has been corrected for in all of
  the following.}.

\begin{figure}
 \resizebox{\hsize}{!}{\includegraphics{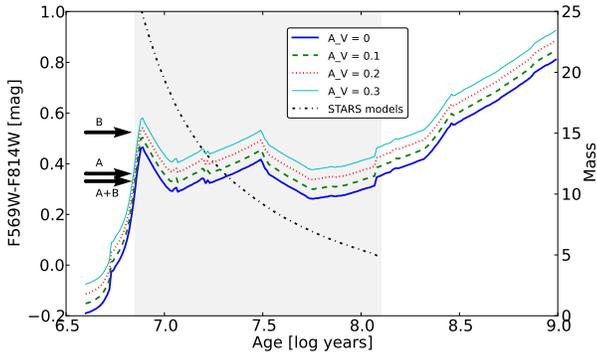}}
  \caption{Padova integrated model $F569W$-$F814W$ colours of a stellar population (on the left y-axis), 
  are plotted against the age of the population for a range of extinctions. The observed colours of 
  regions A and B, together with the total colour of the entire region (A+B) are indicated with arrows, 
  these values have been corrected for Milky Way extinction only. The range of population ages 
  (approximately 10 - 100 Myr) which are consistent with the cluster colours are indicated by the 
  shaded region. Unfortunately there is a strong degeneracy in the $F569W$-$F814W$ colour of the models.
On the right y-axis, the ZAMS mass of single stars from the STARS code are plotted against their lifespan. 
A population age of 10 Myr is consistent with a progenitor mass of 25 \msun{}, while for 
an older population the progenitor must be less massive, likely $\sim$ 8 \msun{}.}
\label{fig:cluster}
\end{figure}

While the SN is not coincident with either of the two complexes A and
B, an association is still plausible.  The SN is $\sim$5 pixels
(0.5\arcsec) from the brightest pixels in A and B. At the distance of
NGC 7479, and scaling by ${4}\over{\pi}$ to account for projection
this corresponds to $\sim 100$ parsec. If we assume a velocity of 100
\kms{} for the progenitor, it could traverse this distance in $\sim$1
Myr. As this is a factor of ten less than the lifespan of a single
25~\msun{} star (Fig.~\ref{fig:cluster}), the progenitor could well 
be associated with either region A or B (or indeed be unrelated to both).

The {\sc hstphot} package \citep{Dolphin00} was used to produce
photometry of all sources detected in the region of the SN\footnote{We
  used the pre-processing programs that accompany {\sc hstphot} to
  remove cosmic rays and hot pixels ({\sc mask, crmask, hotpixels}),
  to combine the individual exposures in each filter ({\sc coadd}) and
  to measure the sky background ({\sc getsky}).}.  {\sc hstphot} was
then run separately for the coadded F569W and F814W filter images with
a detection threshold of 3$\sigma$.

\begin{figure}
 \resizebox{\hsize}{!}{\includegraphics{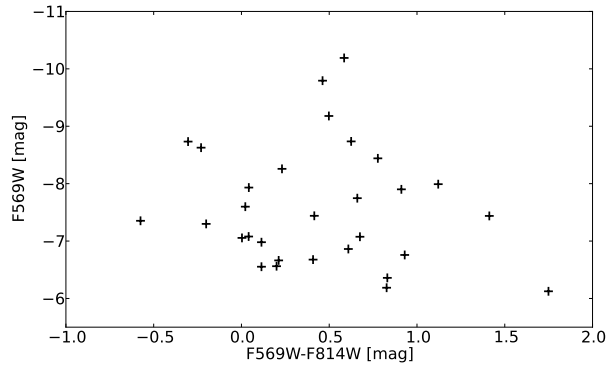}}
  \caption{All sources detected by {\sc hstphot} above the 3$\sigma$
    level in both the $F569W$ and $F814$ filters. Magnitudes have been
    corrected for the distance of NGC 7479 and for Milky Way extinction,
    but not for extinction in the host. Furthermore, we have made no
    attempt to separate out sources which are poorly fit by a PSF, and
    are hence likely unresolved clusters. Indeed, any source with an
    absolute magnitude brighter than $-8.5$ mag is likely a cluster rather than a
    single star.}
   \label{fig:local_pop}
\end{figure}

A colour-magnitude diagram of all sources detected by {\sc hstphot} in
both filters, within $\sim$500 pc of SN 2009jf is shown in
Fig.~\ref{fig:local_pop}. 
As can be seen, there are no evolved red sources detected in both filters, 
and indeed the population appears quite blue, which is indicative of a young and
massive stellar population.
While the population of detected sources appears blue, the limiting
magnitudes of the images likely prevent us from detecting most of the  evolved red
supergiant population, or the main sequence below about 20-30 \msun{}. 
As such it is difficult to provide an estimate of an age for the
surrounding stellar population. It is likely that most objects
brighter than $-8.5$ are unresolved, compact clusters. Furthermore the
region is similar to the 50-300pc sized starforming complexes that are
common in late-type spirals \citep{bastian05}. Such complexes
typically contain compact star clusters of $<10$pc diameter, which are
likely to host coeval populations themselves. However the whole star
forming complex may have a significant age spread 
\citep[see discussion in ][]{Crockett08}. 
The closest source detected by {\sc hstphot} is
66 mas (10 pc) away from the nominal SN position in the F569W
($m_{F569W}$ = 23.75). With an absolute magnitude of $M_{F569W}$
=-9.25, this is likely to be a compact cluster. The host region of
SN2009jf is quite similar to that of SN2007gr \citep{Crockett08},
with the SN being close to (but not exactly coincident) with a compact 
cluster and contained with a larger starforming complex.

We can also attempt to estimate the age of the population by comparing
the observed colours of the region to models. If we can determine the
age of the A and B region, then we can infer the most massive stars 
which would be still extant at that time. We have performed simple 
aperture photometry on the $F569W$ and $F814W$ filter WFPC2 images on the
regions indicated A and B in Fig.~\ref{fig_F569}. Using the revised
zeropoints of Dolphin (2009), and correcting for Milky Way extinction,
we find the colours of regions A and B to be M$_{F569W-F814W}$= 0.36
and 0.52 mag respectively. In a large aperture encompassing both
regions (also indicated in Fig.~\ref{fig_F569} with a dashed circle),
we find a colour of M$_{F569W-F814W}$=0.33 mag, which is reassuring as
this is a similar colour to region A, which contributes most of the
flux. We have no constraint on the internal extinction in the
complexes, so we will regard these colours as upper limits on the true
colours, which are likely bluer.

We have used the Padova stellar population
models\footnote{http://stev.oapd.inaf.it/cmd}
\citep{girardi02,marigo08} to create a table of integrated magnitudes
for a single stellar population, at a metallicity appropriate to the
site of SN 2009jf (Z=0.012), and in the WFPC2 filter system.  The
Padova model colours are shown as a function of age in Fig.~\ref{fig:cluster}, 
together with the lifetimes of massive stars as found from the STARS
evolutionary code \citep{Stancliffe09}.  The degeneracy in the
M$_{F569W-F814W}$ colour versus age plot precludes us from determining
the age for the clusters. Region A is consistent with an unreddened
population of stars with a maximum progenitor mass somewhere between
8 and 25 \msun. 
Complex B is redder which implies an older age. But the 
degeneracy between age and extinction prevents a useful 
determination. For example a moderate amount of internal 
extinction would make the age range consistent with a 
8-25 \msun\ population.  
Unfortunately the compact cluster which is closest to the progenitor 
position is detected only in the $F569W$ filter, and so we cannot 
constrain its age, beyond the fact that it is blue, and hence 
presumably among the younger objects in the region.
As for the progenitor of SN2007gr and its possible host cluster, 
future observations of the region with HST in the $U$ and 
$B$ bands could age date the cluster.

We also present an H$\alpha$ image of NGC 7479, which was obtained on
13 September 1996 with the Prime Focus Cone Unit + CCD on the Isaac
Newton Telescope. This deep image consisted of an 1800 s exposure
taken with the H626 filter (\Ha{}) and another 1800 s exposure taken
with the H712 filter to allow for removal of continuum light.  The
{\sc hotpants} package (A. Becker) was used to subtract the continuum
image from the \Ha{} image. 
The H$\alpha$ flux does not appear to be from a point source, but 
is rather spread out over several pixels at the southern end of the 
complexes. The SN position in the H$\alpha$ image was determined 
by alignment to a Liverpool Telescope $r$'-band image, and is found 
to be on the edge of this region of H$\alpha$ flux (as shown in Figure 
\ref{figha}).

In conclusion, the environment of the SN clearly displays signs of
recent star formation, with a strong \Ha{} flux and integrated colours
which are consistent with a young, massive stellar population. A
colour magnitude diagram of sources within a radius of 500 pc of the
SN indicates a young population, with no detections of the red
supergiants which may be expected in a slightly more evolved
population. The progenitor is close to, but not coincident with, two
regions which we have termed A and B. Unfortunately we cannot
distinguish between a high mass ($\sim$25 \msun) or low mass ($\sim$10
\msun) progenitor on the basis of the age of the clusters due to the
degeneracy in the age-colour relation. However, on the basis of the
H$\alpha$ flux and the surrounding blue stellar population, together
with the characteristics of the SN, we consider the high mass channel
more likely. Future observations after the SN has faded may help to
better address this issue.

\section{Bolometric light curve of SN 2009jf}

\label{parametribolo}

\begin{figure}
\includegraphics[width=9cm,height=7cm]{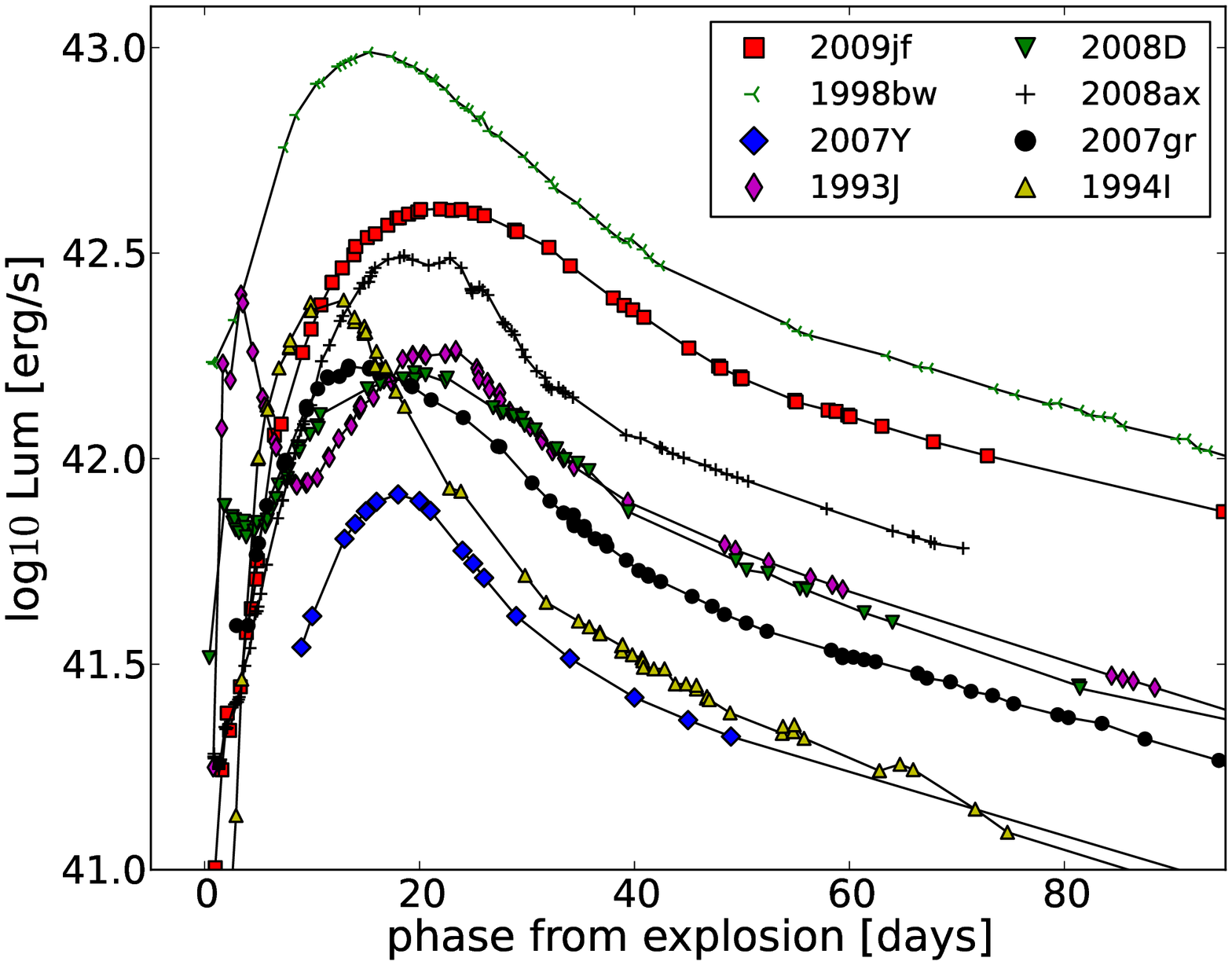}
\caption{The bolometric light curve of SN 2009jf compared with those of
other core-collapse SNe. References for each SN:
- SN 2009jf: this work;
- SN 1998bw: $E(B-V) =$ 0.06 mag, $\mu =$  32.76 mag , phot. data:
\protect\cite{patat01};
- SN 1994I: $E(B-V) =$ 0.04 mag, $\mu =$  29.60 mag
\protect\citep{sauer06}, phot. data: \protect\cite{rich96};
- SN 1993J: $E(B-V) =$ 0.079 mag \protect\citep{barbon95}, $\mu =$ 27.8
mag \protect\citep{Freedman94}, phot data: \protect\cite{barbon95};
- SN 2008D: $E(B-V) = $ 0.65, $\mu =$ 32.29 mag \protect\citep{mazzali08}, 
phot. data: \protect\cite{mazzali08,modjaz09};
- SN 2007gr: $E(B-V) =$  0.092 mag, $\mu =$ 29.84 mag, phot. data:
\protect\cite{valenti08b,hunter09};
- SN 2008ax: $E(B-V) =$  0.40 mag \protect\citep{taubenberger11}, $\mu =$
29.92 mag \protect\citep{pastorello08a}, phot. data:
\protect\cite{pastorello08a,taubenberger11};
- SN 2007Y: $E(B-V) = $ 0.112 mag, $\mu=$ 31.13 mag , phot. data:
\protect\cite{stritzinger09}.
}
\label{figbolo}
\end{figure}

The bolometric light curve of a SN is a powerful tool to validate
model predictions. It is well known that, starting from $\sim$ 1 week
past explosion, the light curve of a non-interacting stripped-envelope
SN is mainly determined by the amount of nickel synthesised during the
explosion, together with the amount of ejected material and the
kinetic energy \citep{arnett80,arnett82}.  As a first-order
approximation to the true bolometric luminosity, we integrated the SN
emission in the spectral window accessible from ground, from the UV
atmospheric cut off to the NIR ($JHK$)\footnote{Photometric data in
  the Sloan filters were transformed from the AB to VEGA system
  \protect\citep{Holberg06}. }.  All magnitudes were then converted to
fluxes\footnote{The zeropoints for the conversion have been computed
  by integrating the spectrum of Vega with the Landolt and Sloan
  filters \protect\citep{buser78,gunn98}.}  and integrated from $U$ to
$K$ using Simpson's Rule. The resulting bolometric light curve is
shown in Fig.~\ref{figbolo} together with the bolometric light curves
of other stripped envelope SNe. SN 2009jf is one of the brightest SNe Ib
observed to date. At 20.5 days after explosion, it reached a maximum
luminosity of log$_{10}$L = 42.62 $\pm$ 0.05 erg s$^{-1}$, in
between the Type Ic GRB-SN 1998bw and SN 2008ax.  The light curve of
SN 2009jf evolves quite slowly, only reaching the radioactive tail
around 50 days after explosion. The slope of the tail is then 0.0133
mag day$^{-1}$, which is faster than the cobalt decay time scale.

\cite{valenti08a} developed a toy-model in order to compute a
first-order estimate of the main physical parameters that shape the
bolometric light curve \citep[see also][]{chatzopopoulos09}:
$M_{\mathrm{56Ni}}$, $M_{\mathrm{ej}}$ and $E_{\mathrm{k}}$. The toy-model is based on very
simple approximations \citep{arnett82,cappellaro97,clocchiatti97},
dividing the light curve into an optically thick (photospheric) and an
optically thin (nebular) phase.  Adopting a photospheric velocity at
maximum of 11000 \kms{} and an optical opacity of $k_{opt}$= 0.06, we
obtained the following parameters: $M_{\mathrm{56Ni}}$ = 0.23 $\pm$ 0.02
\msun{}, $M_{\mathrm{ej}} = 5-7$ \msun{} and $E_{\mathrm{k}} = 5.5-10.6 ~ 10^{51}$ erg.
In Appendix~\ref{ap1} we report in more detail on the toy model as
applied to the case of SN 2009jf and other SNe Ib, along with some
caveats.  The ejected mass we find for SN 2009jf is comparable with
those obtained for other massive SNe Ib (SN 2008D,
\citealt{mazzali08}, \citealt{tanaka09}; SN 1999dn,
\citealt{benetti11}) and for some massive and energetic SNe Ic.  The
kinetic energy obtained is quite high. This is not surprising however,
since our toy-model also seems to overestimate the kinetic energy for
other SNe Ib.  The true kinetic energy of SN 2009jf is likely close to
the lower edge of our range. Nevertheless the values we find for the
kinetic energy and ejected mass are consistent with those of
\cite{Sahu11}. The ejected nickel mass of \citeauthor{Sahu11} (0.17
$\pm$ 0.03 \msun) is slightly smaller than our value, most likely
because they did not include the infrared contribution in their
bolometric light curve\footnote{Their over-estimate of the radioactive
  tail magnitude does not change their nickel mass estimate as this is
  mainly based on the SN magnitude at maximum light.}.

\section{Discussion}

In the last 5 years, good data-sets have been published for several
SNe Ib (SN 2007Y, \citealt{stritzinger09}; SN 2008D,
\citealt{Soderberg08,mazzali08,modjaz09,malesani09}; SN 2008ax,
\citealt{pastorello08a,Chornock10,taubenberger11}; SN 2009jf,
\citealt{Sahu11}, this work).  These data allow us for the first time
to make comparisons between the different stripped-envelope SNe
subtypes.  The broad light curves and the evidence that some SNe Ib
are quite luminous and energetic, suggests that these explosions do
not always release the canonical 10$^{51}$ erg, as often assumed.

In fact, SN 2007Y, one of the faintest and least massive He-rich SNe
\citep{stritzinger09}, ejected $\sim$ 2 \msun{} of material once the
He (1.5 \msun) and residual H (0.1 \msun) layers \citep{maurer10}, are
taken into account\footnote{\protect\cite{stritzinger09} found that
  excluding He and H, the ejected mass of SN 2007Y was $\sim$ 0.45
  \msun.}.  In this context, SN 2009jf is one of the most energetic
and massive SNe Ib known to date.

\begin{figure}
\includegraphics[width=9cm,height=7cm]{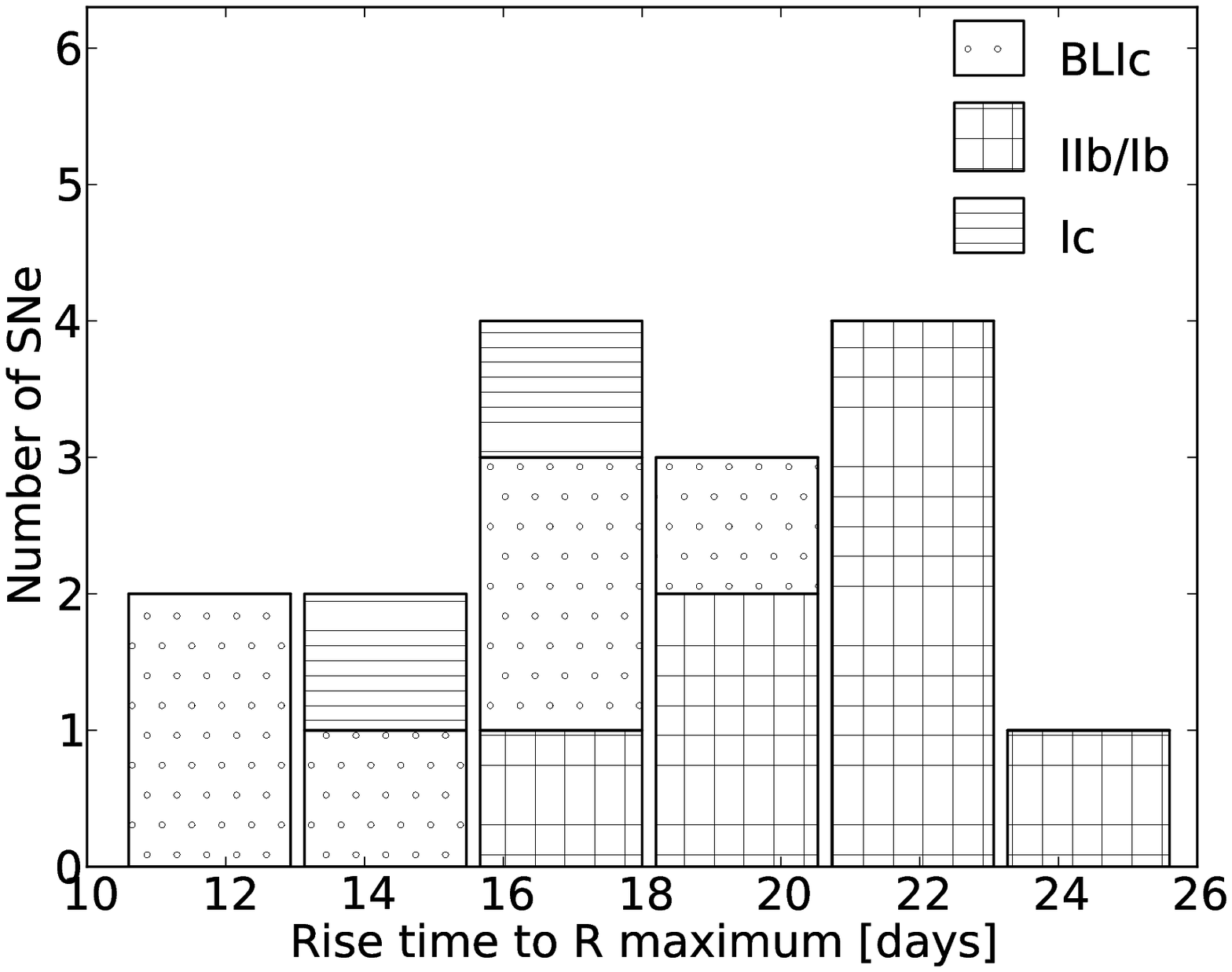}
\caption{The rise times of a sample of stripped-envelope
  SNe. References for each SN: 
  -1998bw: \protect\cite[from GRB 980425, ][]{GRB980425}; 
  -1994I: \protect\cite[modelling, ][]{sauer06}; 
 -2003jd: \protect\cite[pre-explosion limit][]{valenti08a}; 
 -1999ex: \protect\cite[early discovery, ][]{stritzinger02}; 
  -2002ap: \protect\cite[modelling, ][]{mazzali02a}; 
  -1997ef: \protect\cite[modelling, ][]{mazzali00a}; 
  -2006aj: \protect\cite[from GRB060218, ][]{cusumano06};
  -1993J: \protect\cite[early discovery, ][]{lewis94}; 
  -2007gr: \protect\cite[good pre-explosion limit, ][]{valenti08b}; 
  -2008D: \protect\cite[from swift detection, ][]{berger08}; 
  -2008ax: \protect\cite[pre-discovery limit, ][]{pastorello08a}; 
 -1996cb: \protect\cite[1 day from early discovery][]{Qiu99}; 
 -1983N: \protect\cite[1 day from early discovery, ][]{Richtler83}; 
  -2007Y: \protect\cite[1 day from early discovery, ][]{stritzinger09};
  -2009bb: \protect\cite[good pre-explosion limit, ][]{pignata11}.}
\label{figrise}
\end{figure}

From Fig.~\ref{figbolo}, it is clear that, while some He-poor SNe
(eg. SNe 1998bw, 2007gr) show an asymmetric peak (with a faster rise
than decay), all the He-rich SNe show a slow rise to maximum.  In
Fig.~\ref{figrise} we show the rise times for a sample of
stripped-envelope SNe. In the sample of published SNe Ib we have tried
to include only SNe with reasonable information on the explosion epoch
(early discovery or good pre-discovery limit) and good data coverage.
SN 1983N has also been included because of the early $V$-band
discovery, even though the light curve coverage is not ideal.
Only two SNe Ic, with data published, have been monitored soon after
the explosion, although for some broad-lined SNe Ic the explosion
epoch is well constrained by the associated X-ray flash or Gamma Ray
Burst. We have excluded stripped-envelope SNe for which there is
evidence of circumstellar interaction, as interaction will increase
the luminosity of the supernova.  We also excluded the new class of
ultra-bright SNe \citep{quimby09,pastorello10} that share some
similarities with SNe Ic \citep{pastorello10}, for which a flux
contribution to the light curve due to interaction has also been
proposed \citep{Blinnikov10}.  These SNe show rise times of up to
$\sim$100 days, and may have an origin different to \emph{normal} Ic
SNe.

Even though the longer rise times for He rich SNe are clearly apparent, we must 
consider the possibility that this is a selection effect. SNe Ib are quite rare, and 
slowly-evolving/rising SNe Ib are easier to discover at an earlier phase than fast 
rising SNe Ib. However, our findings are confirmed by \cite{Drout10} who recently 
presented an extended set of stripped-envelope light curves in which there are 
no fast rising SNe Ib, and SN 1994I is the only fast evolving stripped-envelope SN. 

Recently, several fast evolving stripped-envelope SNe have been
discovered (or recovered in archival data), but unfortunately none of
them has been well covered in the premaximum phase (e.g. SN 2008ha,
\citealt{valenti09}, \citealt{foley09}; SN 2005E \citealt{Perets10};
SN 2002bj, \citealt{Poznanski10}; SN 2005cz, \citealt{Kawabata10}).
It is also highly debated whether they are core-collapse SNe,
thermonuclear explosions in a \emph{non-canonical} Ia scenario, or
even new explosion channels.
If we believe that these fast evolving stripped-envelope SNe are not
core-collapse explosions, we will conclude that, with the exception of
SN 1994I, all stripped-envelope SNe ejected at least 2 \msun{} of
material.

The large ejected mass (3-7 \msun) and large oxygen 
mass (the latter reported by Sahu et al. 2011) of SN2009jf
could be explained by the explosion of a 5-9 \msun{} CO star (assuming
1.5 \msun neutron star remnant).  The WC population in the 
LMC  have pre-SN masses between 6-18 \msun{} \citep{Crowther02}
A 5-9 \msun CO star could originate in a single massive progenitor 
of $M_{ZAMS}$ $>$ 35 \msun{} \citep{Georgy09,Eldridge04}
with radiative driven mass loss, suggesting that a massive WR star
(with a low He residual mass) is plausible.  Equally a lower mass star 
which has been stripped of its envelope through binary transfer is 
also possible - the CO core mass of a 20 \msun star is about 5 \msun
in the STARS models \citep{Eldridge04}.  The analysis of the 
progenitor environment cannot distinguish between these two, as the 
age of the stellar population in the vicinity of the SN progenitor is 
not constrained well enough to determine a star formation age. 
The fact that the SN is very close to a compact star cluster suggests
that an age determination of that cluster could provide further
constraints, if UV data are obtained once the SN has faded.

The detection of relatively weak He at high velocity and the spectral
similarity with the He poor SN 2007gr suggest the presence of only a
small amount of He in the ejecta of SN 2009jf.  However, this
conclusion needs to be confirmed by detailed spectral modelling of SN
2009jf.

The profile of [\OI] \LL{}6300,6364 is suggestive of an asymmetric
explosion with an off-axis dense core with clumps. Similar
conclusions can be drawn for the magnesium and calcium distribution by
comparing their line profiles to that of oxygen.  While a similar
profile is expected for \MgI{}] and \OI{} as they come from similar
  regions in the ejecta, [\CaII] \LL{}7291,7324 should form in the
  inner part of the ejecta where silicon is more abundant, or in the
  outer layer of He \citep{fransson89}. The similar profile for the
  three lines could indicate that mixing is an important factor in the
  explosion of SN 2009jf.

\section{Summary}

In this paper we have presented optical and infrared photometry and
spectroscopy of SN 2009jf spanning from $\sim$ 20 days before $B$-band
maximum to one year after maximum.

We have shown that SN 2009jf is a slowly evolving, massive and
energetic stripped-envelope SN which retained only a small part of its
He layer at the moment of explosion. The SN exploded in a young
stellar environment, and the progenitor is likely a massive star $>
25-30$ \msun{} as suggested by \cite{Sahu11}.  Furthermore, the
similarity with the SN Ic 2007gr suggests a similar progenitor for at
least some SNe Ib and Ic. The nebular spectra of SN 2009jf are
consistent with an asymmetric explosion with an off-center dense core.

We have also shown that He rich SNe appear to have longer rise times
than other stripped-envelope SNe. However, this should be treated as a
preliminary result, and needs to be verified with a larger sample of
stripped-envelope SNe, while carefully accounting for all possible
systematic biases.

\section*{Acknowledgements}
We thank the anonymous referee for helpful suggestions. 
S.V. is grateful H. Wang for hospitality at the UCLA.
G.P. acknowledges support by the Proyecto FONDECYT 11090421.  S.B.,
E.C., M.T.B., F.B., P.A.M. and M.T. are partially supported by the
PRIN-INAF 2009 with the project \emph{Supernovae Variety and
  Nucleosynthesis Yields}.  G.P., M.H. and J.M. acknowledge support
from the Millennium Center for Supernova Science through grant
P06-045-F funded by \"{}Programa Bicentenario de Ciencia y Tecnolo\'gia de CONICYT\"{}, 
\"{}Programa Iniciativa Cientifica Milenio de MIDEPLAN\"{},
from Centro de Astrofi�sica FONDAP 15010003 and by Fondecyt through grant
1060808 from the Center of Excellence in Astrophysics and Associated
Technologies (PFB 06).  S.T. acknowledges support by the TRR 33 "The
Dark Universe" of the German Research Foundation.
S.M. acknowledges support from the Academy of Finland (project 8120503)
E.K. acknowledges financial support from the Finnish Academy of Science 
and Letters (Vilho, Yrj\"{o} and Kalle V\"{a}is\"{a}l\"{a} Foundation).
This paper is based on observations made with the following
facilities: ESO Telescopes at the La Silla and Paranal Observatories
under programme IDs 184.D-1151, 085.D-0750 and 386.D-0126, the Italian
National Telescope Galileo (La Palma), the 1.82~m Copernico telescope
of the Asiago Observatory (Italy), the William Herschel (La Palma),
Liverpool Telescope (La Palma), Nordic Optical Telescope (La Palma),
AlbaNova Telescope (Sweden), Prompt telescopes (Chile), Calar Alto
(Spain).
We thank Genoveva Micheva for help with observations.
We are grateful to the staff at all of  the telescopes for their assistance. 
This paper makes use of data obtained from the Isaac Newton Group
Archive which is maintained as part of the CASU Astronomical Data
Centre at the Institute of Astronomy, Cambridge.  Based on
observations made with the NASA/ESA Hubble Space Telescope, obtained
from the data archive at the Space Telescope Institute. STScI is
operated by the association of Universities for Research in Astronomy,
Inc. under the NASA contract NAS 5-26555.
We thank K. Itagaki for providing us his images of SN 2009jf.
This manuscript made use of information contained in the Bright
Supernova web pages (maintained by the priceless work of D. Bishop),
as part of the Rochester Academy of Sciences
(http://www.RochesterAstronomy.org/snimages).
This publication makes use of data products from the Two Micron All
Sky Survey, which is a joint project of the University of
Massachusetts and the Infrared Processing and Analysis
Center/California Institute of Technology, funded by the National
Aeronautics and Space Administration and the National Science
Foundation.

\appendix

\section[]{Data reduction}
\label{ap0}

The \emph{Swift} data were reduced using the \emph{Swift}
pipeline and colour corrected using the recipe of \cite{li06}. The
data from the Prompt telescopes were pre-reduced by the MCSS group,
while for those from the robotic Liverpool+RATCAM telescope we used the
dedicated pipeline. All the other data were spectroscopically and
photometrically pre-reduced with the {\sc quba} pipeline\footnote{The
  {\sc quba} pipeline was written by S. V. for the reduction and
  analysis of spectro-photometric observations of SNe with the
  instruments and configurations commonly used within the ELP
  collaboration.}. The pre-reduction in the {\sc quba} pipeline (bias,
trim and flat field correction) is performed using {\sc python pyraf}
scripts.  The magnitude is measured with a PSF-fitting technique
(using {\sc daophot}). Template subtraction is also implemented using
{\sc sregister} package within {\sc iraf} to register the images and
{\sc isis mrjphot}\footnote{{\sc isis} is an image subtraction package
  \citep{alard1,alard2}, {\sc mrjphot} is the image subtraction task
  within {\sc isis}.} to subtract an image without the SN to the images with the SN.
In the case of SN 2009jf, the SN magnitudes have been measured without
using template subtraction for all the data from discovery to 3 weeks
after the $B$-band maximum. For these epochs the SN was bright enough
to neglect the contamination from the close-by clusters. The rest of
the data were measured after the subtraction of archival template
images acquired prior to the SN explosion. A comparison of the SN
magnitude as measured with and without template subtraction was made
(when the SN was bright) to ensure both methods are consistent.  The
contamination from the close-by clusters can be neglected until 5
weeks after maximum. After that, the SN magnitude is over-estimated by
amounts ranging from 0.01 mag in the redder bands at 5 weeks after
maximum, to 0.5 mag in $B$ at 6 months post maximum, 
if no template subtraction is performed. Since the
contamination is larger in the blue bands, SN magnitudes in the $U$-
and $u$-band data were measured after template subtraction at all
epochs. In the $z$ band no good templates were found in the archive,
and so PSF-fitting was used also for the late data points. However,
the contamination from the bright clusters should be less than 0.1 
mag for our faintest measurement in $z$, and even smaller in the previous epochs.

We found a discrepancy between the \emph{Swift} $U$-band photometry
and the rest of our $U$-band data. The \emph{Swift} $U$-band
photometry using the \emph{Swift} pipeline, colour corrected using
the recipe of \cite{li06}, is initially fainter than our photometry by
$\sim$ 0.1 mag, but becomes brighter by $\sim$ 0.2 mag as the SN
fades. Due to the contamination from the nearby cluster, we re-reduced
the \emph{Swift} photometry using the template subtraction technique
and applied an S-correction \citep{pignata08} as the \emph{Swift}-$U$
band filter is bluer than the $U$ passband in the Landolt
system. After making these corrections, the late \emph{Swift}-$U$ band
photometry is more consistent with data from other telescopes,
although still fainter by $\sim$ 0.1 mag at maximum light. The strong
contamination from the cluster in the $U$ band also gives us cause for
concern for the $uvw1$ \emph{Swift} data, given the red tail of
the \emph{Swift} $uvw1$ passbands \citep{brown10}. The swift data in
the $uvm2$ and $uvw2$ filters, as reduced with the \emph{Swift}
pipeline show an almost constant magnitude, which is likely due to the
bright cluster and not to the SN.  For this reason the $uvm2$ and
$uvw2$ magnitudes are not reported in the Tables.

Observations of standard fields \citep{landolt92,smith02} during
photometric nights were used to calibrate the magnitudes of a local
sequence of reference stars (see Fig. \ref{figseqstar} and Tables
\ref{tabseqstar1} and \ref{tabseqstar2}), which were used to calibrate
the SN magnitudes obtained under non-photometric conditions. The
apparent magnitude of the SN was computed using the zero point from
the local sequence stars, together with a first order colour
correction. The unfiltered images provided by amateur astronomers have
been calibrated taking into account the CCD response as a function of
wavelength. Some images were acquired with a Bayer filter; these were
reduced and calibrated following \cite{hoot07} to $BVR$ filters in the
Landolt system applying only a zero point.

The $JKH$ data have been calibrated using the bright stars from Two
Micron All-Sky Survey (2MASS) in the field of SN 2009jf and reported
in Table \ref{tabinf}. The $UBV$ and $uvw1$ \emph{Swift} data are
reported in \ref{tabswift}. The $uvw1$ band were calibrated to the
\emph{Swift} system magnitudes.

The spectroscopic data were reduced using the {\sc quba} spectra
reduction tasks, which are based on standard spectroscopic {\sc iraf}
tasks. A check on the wavelength calibration is performed using the
sky lines, and second order corrections are made to the
NOT+ALFOSC+Grism4 spectra \citep{vallery07}. The final step in
spectral reductions with the {\sc quba} pipeline is the removal of
telluric lines, using the standard stars observed in the same
night. The fluxes of the reduced spectra were compared with photometry
of the SN at the same epoch, and, when necessary, multiplied by a
constant to correct for slit losses.

The NaCo images were reduced in the standard manner using the {\sc
  xdimsum} package within {\sc iraf}.  The sky was subtracted using
sky frames created from median-combined dithered on-source images.

The HST images ($F569W$ and $F814W$ filters aquired on 1995 October
16th) were obtained from the ESO archive, and reduced with
the OFT (On-The-Fly) pipeline, which applies the most up-to-date
calibrations available at the time of request. Multiple exposures were
taken in each filter to facilitate the rejection of cosmic rays. These
images were combined using the {\sc crrej} task within the {\sc iraf
  stsdas} package. The site of the SN fell upon the WF4 chip, which
has a pixel scale of 0.1\arcsec per pixel.

\section[]{Toy Model and Ib SNe}
\label{ap1}

\cite{valenti08a} introduced a toy model to compute a first-order estimate 
of the main physical parameters $M_{\mathrm{56Ni}}$, $M_{\mathrm{ej}}$, $E_{\mathrm{k}}$. 
Here, we report the application
of this toy model to 5 well studied He-rich SNe (2008D, 1999dn,
2008ax, 2007Y and 2009jf) and discuss some caveats.

The photospheric velocities at maximum \citep[used as scale velocity,
  see][]{arnett82} are reported in Table \ref{tabmodel}, together with
the values obtained from the model fit and values from the
literature. The Nickel masses obtained with our toy model are
consistent with the values from more elaborate models. The ejecta
masses for SNe 2008ax and 2008D are also in agreement with other
estimates. On the other hand, the kinetic energies obtained for SNe
2008D and 2008ax are larger than those obtained by detailed models.

\begin{table*}
 \centering
\begin{minipage}{180mm}
\caption{Physical parameters for stripped-envelope SNe derived from models of the
bolometric light curves$^a$.}  
\label{tabmodel}
\begin{tabular}{@{}cccccccccc@{}}
\hline
SN  & $v_{ph}^b$ &$M_{\mathrm{56Ni}}$(tot) & $M_{\mathrm{ej}}$(tot) & $E_{\mathrm{k}}$(tot)     & $E_{\mathrm{k}}$(inner)   & $M_{\mathrm{56Ni}}$(inner)/ & $M_{\mathrm{56Ni}}(*)$ & $M_{\mathrm{ej}}$(*) & $E_{\mathrm{k}}(*)$  \\
      & (\kms{})      & ($M_{\odot}$)  & ($M_{\odot}$) &($10^{51}$ erg)&($10^{51}$ erg) & $M_{\mathrm{56Ni}}(tot)$   & ($M_{\odot}$) & ($M_{\odot}$) & ($10^{51}$ erg) \\
\hline		    
2009jf   & 11000   &  0.23 $\pm$0.02 &   5-7    &5.5 - 10.6& 0.21 - 0.23  & 0.32 - 0.42    &    --           &      --      & -- \\
1999dn & 10100   &  0.11 $\pm$0.02 &   4-6    & 5 - 7.5    & 0.07 - 0.11 & 0.3 - 0.4       &     --          &      --      &  -- \\
2008D  & 11000    &  0.09 $\pm$0.02 &   5-7    & 8 - 13    & 0.12 - 0.18   & 0.25 - 0.31    & 0.07          & 4.3 - 6.3    &  3.5 - 8.5 \\
2008ax & 10000   &  0.12 $\pm$ 0.02 &   2-4    &  2 - 4      & 0.16 - 0.18 &  0.35 - 0.45   & 0.07 - 0.15 & 1.9 - 4.0   &  0.7 - 2.1 \\
2007Y   &  9000    &  0.04 $\pm$0.01 &  1 -2 & 0.7 - 1.5  & 0.04 - 0.08   & 0.35 - 0.45 & 0.06           & $\sim$ 2 $^{c}$ & $\sim$ 0.82$^{d}$ \\
\hline
\end{tabular} 

$^a$The $E(B-V)$ values and the distance moduli  are
reported in the caption of Fig. \ref{figbolo}. For comparison, we report in the last
three columns [labeled with (*)] the values of $M_{\mathrm{ej}}$, $M_{\mathrm{56Ni}}$, and
$E_{\mathrm{k}}$ computed  with sophisticated spectral and light-curve  modelling (SN 2008ax: \cite{maurer10}, 
SN 2007Y: \cite{stritzinger09,maurer10}, and SN 2008D: \cite{tanaka09}.\\
$^b v_{ph}$ is the photospheric velocity  close to the maximum of the  bolometric light curve used as \emph{scale velocity} \citep{arnett82} .\\
$^c$ This is the total ejected mass including the He and H contribution taken from \cite{stritzinger09} and \cite{maurer10}.\\
$^d$ This is the total energy obtained by \citep{stritzinger09} considering the total ejected mass (H and He included). 
\end{minipage}
\end{table*}

\begin{figure}
 \includegraphics[width=9cm,height=8cm]{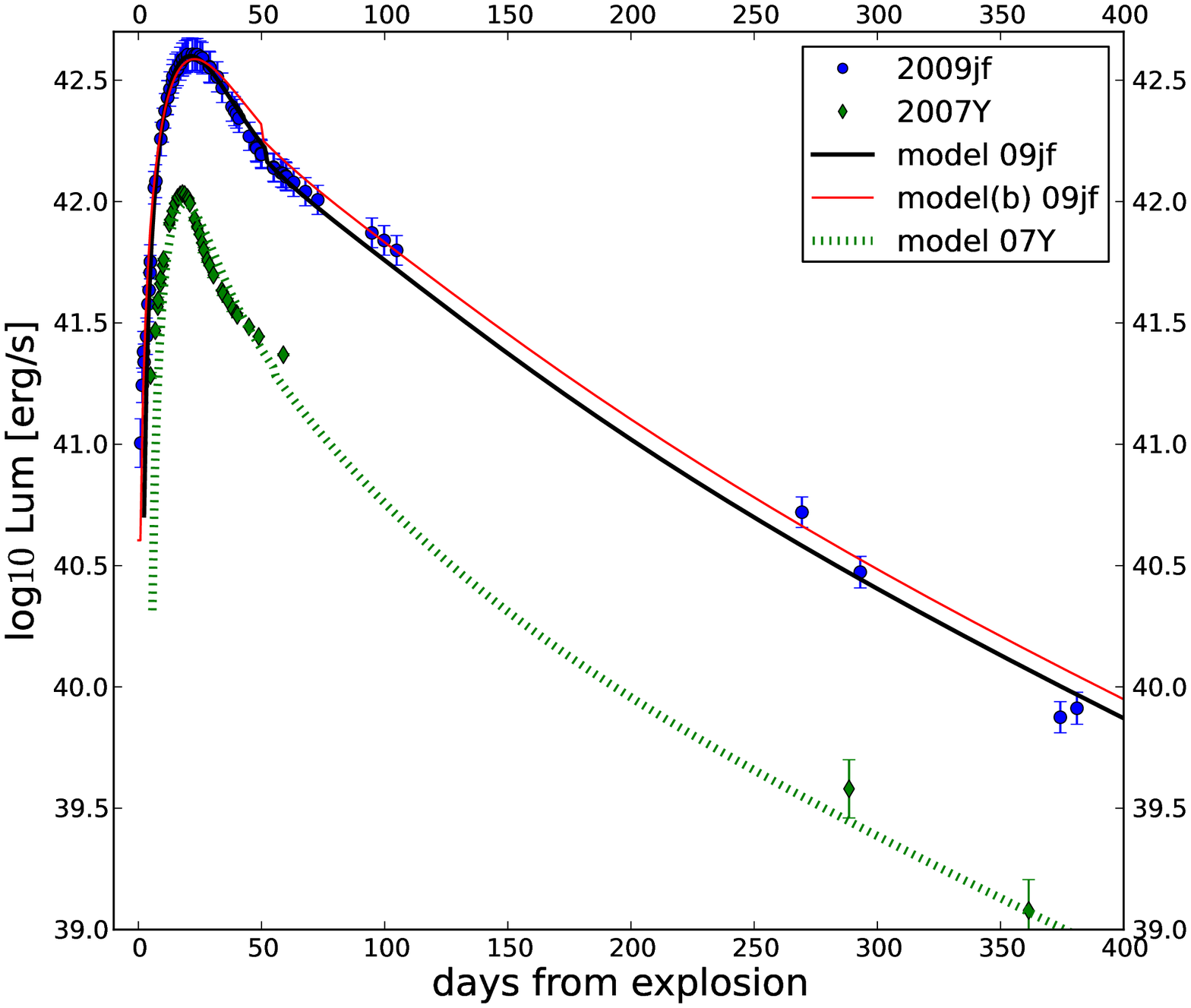}
 \caption{The bolometric light curves of SNe 2009jf and 2007Y and the
 toy models that fit the light curves. The toy model labelled with
 (b) was obtained using also the points soon after the
 explosion. This model can not reproduce nicely the peak, 
 overestimating the ejected mass and kinetic energy.}
  \label{figmodel2}
\end{figure}

We stress that there are caveats inherent in this toy model.  The
model uses a constant opacity \citep[we used $k_{opt}$= 0.06 as in
][]{valenti08a}, while in reality the opacity changes as the SN
expands. Using a larger value for the optical opacity results in a
lower ejected mass and kinetic energy \citep[see also
][]{chatzopopoulos09}. The model also uses a single velocity
(\emph{scale velocity}) that to first order can be approximated as the
photospheric velocity at maximum \citep{arnett82}.  Since no density
structure is considered, the toy model will not distinguish among objects 
with similar photospheric velocities and light curves, but
different density profiles.  The last important caveat concerns the
time interval during which the optically thick (photospheric) and
optically thin (nebular) approximations are valid. For slowly
evolving SNe (e.g. SN 2009jf), the ejecta can be safely considered
optically thick until 40 days after $B$-band maximum, while for the
fast evolving SNe 1994I or 2007Y, the SN is no longer completely optically thick
$\sim$20 days after maximum. We also note that the model fit is poor
soon after explosion, probably because the point-source
approximation\footnote{All the Nickel concentrated in one point at the
  center of the SN.} is not valid at this time.  Thus we usually fit
the model bolometric curve (in the photospheric phase) from $15$ to
$10$ days before maximum to 20 to 40 days after maximum (depending on
the speed of evolution of the SN), as shown in Fig. \ref{figmodel2}).

\section[]{Tables}

\begin{table}
  \caption{Optical photometry of SN 2009jf reference stars (Vega magnitudes in  Landolt system)$^a$}
  \label{tabseqstar1}
  \tiny
  \begin{tabular}{@{}cccccc@{}}
  \hline
Id & $U$ & $B$ & $V$ & $R$ & $I$ \\
\hline
 1  &        --           &  16.054  (014)  & 15.275  (003) &  14.844  (002)  & 14.427  (014) \\
 2  &  17.052  (06)  &  16.717  (026)  & 15.881  (014) &  15.406  (039)  & 14.947  (011) \\
 3  &        --           &  16.712  (024)  & 16.029  (023) &  15.602  (017)  & 15.209  (023) \\
 4  &        --           &  15.930  (030)  & 15.139  (013) &  14.726  (016)  & 14.332  (020) \\
 5  &  18.805  (13)  &  18.062  (043)  & 17.048  (020) &  16.420  (013)  & 15.894  (023) \\
 6  &  17.890  (15)  &  17.654  (017)  & 16.812  (012) &  16.280  (014)  & 15.802  (018) \\
 7  &  17.186  (10)  &  17.196  (012)  & 16.509  (022) &  16.124  (017)  & 15.701  (014) \\
 8  &       --            &  18.664  (030)  & 17.088  (062) &  16.468  (051)  & 15.801  (020) \\
 9	 &       --            &  15.854  (030)  & 14.848  (020) &  14.340  (003)  & 13.841  (004) \\
10 &       --            &  14.736  (019)  & 14.053  (011) &  13.648  (020)  & 13.235  (020) \\
11 &       --            &  18.618  (034)  & 17.345  (044) &  16.460  (006)  & 15.600  (020) \\
12 &       --            &  15.376  (020)  & 14.620  (020) &  14.245  (020)  & 13.820  (020) \\
13 &  19.051  (23)  &  18.022  (038)  & 16.973  (049) &  16.352  (020)  & 15.821  (028) \\
14 &  18.577  (16)  &  17.385  (029)  & 16.185  (015) &  15.424  (024)  & 14.826  (008) \\
15 &       --            &  16.715  (030)  & 15.441  (012) &  14.750  (020)  & 14.163  (020) \\
16 &       --            &  18.419  (050)  & 16.881  (031) &  16.097  (016)  & 15.309  (018) \\
17 &  15.966  (07)  &  15.371  (011)  & 14.457  (009) &  13.947  (021)  & 13.530  (025) \\
19 &  14.556  (02)  &  14.303  (011)  & 13.421  (019) &  12.926  (012)  & 12.425  (010) \\
20 &  17.755  (06)  &  16.991  (029)  & 15.945  (020) &  15.285  (018)  & 14.699  (017) \\
\hline
\end{tabular}
$^a$The uncertainties are the standard deviation of the mean
 of the selected measurements.
\end{table}

\begin{table}
  \caption{Optical photometry of SN 2009jf reference stars (AB magnitudes in Sloan system)$^a$.}
  \label{tabseqstar2}
  \tiny
  \begin{tabular}{@{}cccccc@{}}
  \hline
Id & $u$ & $g$ & $r$ & $i$ & $z$ \\
\hline 
  2   & 17.880  (065)  & 16.286  (023)  & 15.634  (027)  & 15.399  (019)  &  15.300  (011)  \\
  5   & 19.473  (105)  & 17.591  (028)  & 16.688  (027)  & 16.360  (022)  &  16.210  (031)  \\
  6   & 18.781  (047)  & 17.247  (029)  & 16.539  (018)  & 16.251  (014)  &  16.150  (040)  \\
  7   & 18.072  (055)  & 16.871  (031)  & 16.337  (017)  & 16.129  (023)  &  16.069  (034)  \\
13   &        --            & 17.544  (039)  & 16.614  (023)  & 16.274  (024)  &  16.053  (026)  \\
14   &        --            & 16.795  (014)  & 15.711  (040)  & 15.303  (016)  &  15.092  (032)  \\
17   & 16.680  (038) & 14.917  (029)  & 14.189  (023)  & 13.975  (022)  &  13.906  (025)  \\
18   &         --           &          --          & 18.033  (058)  & 16.560  (055)  &  15.849  (029)  \\
19   & 15.408  (023)  & 13.834  (010)  & 13.154  (015)  & 12.863  (020)  &  12.727  (020)  \\
20   & 18.596  (055)  & 16.472  (026)  & 15.563  (021)  & 15.181  (022)  &  14.981  (022)  \\
\hline
\end{tabular}
$^a$The uncertainties are the standard deviation of the mean
 of the selected measurements.
\end{table}

\begin{table*}
 \centering
 \begin{minipage}{160mm}
  \caption{Optical photometry of SN 2009jf (Vega magnitudes in Landolt system)$^a$.}
  \label{tablandolt}
  \scriptsize
  \begin{tabular}{@{}ccccccccc@{}}
  \hline
   Date & JD $-$ & Phase$^{b}$& $U$ & $B$ & $V$ & $R$ & $I$ & Source$^{c}$\\
        & 2,400,000\\
\hline
 2009-09-23  &   55097.51  & $-21.9$ &    $-$              &    $>$ 19.5     &   $>$      19.0   &  $>$    18.5     &   $-$                &  NAR \\
 2009-09-24  &   55098.51  & $-20.0$ &    $-$              &    $>$ 19.5     &   $>$     19.0    &  $>$   18.6      &   $-$                &  NAR \\
 2009-09-26  &   55100.50  & $-18.9$ &    $-$              & 19.12 0.50      &   18.90 0.56     &  18.42  0.32     &   $-$                &  NAR \\
 2009-09-26  &   55101.11  & $-18.3$ &    $-$              &    $-$              &    $-$               &   17.91             &   $-$                &  IAUC \\
 2009-09-27  &   55101.61  & $-17.8$ &    $-$              & 18.30   0.600  &  17.92   0.39    &  17.63  0.23     &   $-$                &  NAR \\
 2009-09-29  &   55103.58  & $-15.8$ &    $-$              &    $-$              &  17.36   0.20    &   $-$                &   $-$                &  RCOS20 \\
 2009-09-30  &   55104.58  & $-14.8$ &    $-$              &    $-$              &  17.00   0.10    &   $-$                &   $-$                &  RCOS20 \\
 2009-10-01  &   55105.60  & $-13.8$ &    $-$              &    $-$              &  16.55   0.10    &   $-$                &   $-$                &  RCOS20 \\
 2009-10-01  &   55106.00  & $-13.4$ &    $-$              &    $-$              &  16.376 0.073  &  16.136 0.079  &  16.030 0.077  &  PROMPT  \\
 2009-10-02  &   55106.63  & $-12.8$ & 16.910 0.028  & 16.917 0.012  &  16.359 0.010  &  16.138 0.007  &  16.016 0.011  &  TNG  \\
 2009-10-02  &   55106.64  & $-12.8$ &   $-$               & 16.927 0.041  &    $-$               &    $-$               &    $-$               &  PROMPT \\
 2009-10-04  &   55108.62  & $-10.8$ &   $-$               &    $-$              &    $-$               &  15.693 0.010  &    $-$               &  FORS2 \\
 2009-10-04  &   55109.41  & $-10.0$ & 16.200 0.202  & 16.310 0.015  &  15.805 0.011  &  15.550 0.014  &  15.424 0.013  &  LT  \\
 2009-10-05  &   55109.61  & $ -9.8$  &    $-$              &        $-$          &  15.65   0.10    &      $-$             &   $-$                &  RCOS20 \\
 2009-10-05  &   55110.33  & $ -9.1$  &      $-$            & 15.97  0.12     &  15.66   0.15    &  15.45   0.18    &  15.32  0.11     &  AN1  \\
 2009-10-06  &   55111.36  & $ -8.0$  & 15.864 0.012  & 15.888 0.017  &  15.538 0.016  &  15.267 0.018  &  15.268 0.031  &  CA \\
 2009-10-06  &   55111.38  & $ -8.0$  & 15.870 0.029  & 15.947 0.009  &  15.510 0.009  &  15.333 0.018  &  15.214 0.011  &  LT \\
 2009-10-07  &   55112.31  & $ -7.1$  &      $-$            & 15.90   0.15    &  15.36   0.12    &  15.17   0.16    &  15.13 0.16      &  AN1 \\
 2009-10-08  &   55112.57  & $ -6.8$  &   $-$               & 15.848 0.017  &    $-$               &   $-$                &    $-$               &  PROMPT  \\
 2009-10-08  &   55113.39  & $ -6.0$  & 15.800 0.018  & 15.791 0.010  &  15.308 0.007  &  15.155 0.009  &  14.988 0.009  &  LT  \\
 2009-10-09  &   55113.56  & $ -5.8$  &   $-$               & 15.724 0.015  &  15.292 0.015  &  15.170 0.014  &  14.958 0.018  &  PROMPT  \\
 2009-10-10  &   55114.64  & $ -4.8$  &   $-$               &   $-$               &    $-$               &  14.98   0.10    &  14.90   0.11    &  PROMPT \\
 2009-10-10  &   55115.39  & $ -4.0$  & 15.749 0.021 & 15.681 0.007  &  15.196 0.007  &  15.015 0.010  &  14.879 0.008  &  LT \\
 2009-10-12  &   55116.55  & $ -2.8$  &   $-$              & 15.592 0.015  &  15.143 0.011  &  14.965 0.010  &  14.803 0.009  &  PROMPT \\
 2009-10-12  &   55117.40  & $ -2.0$  & 15.699 0.018 & 15.570 0.018  &  15.111 0.008  &  14.920 0.007  &  14.737 0.007  &  LT \\
 2009-10-13  &   55117.63  & $ -1.8$  &   $-$              & 15.602 0.027  &  15.103 0.044  &  14.942 0.053  &  14.707 0.055  &  PROMPT \\
 2009-10-13  &   55118.48  & $ -0.9$ & 15.670 0.058  & 15.584 0.017  &  15.056 0.015  &  14.928 0.020  &  14.665 0.035  &  NOT \\
 2009-10-14  &   55119.34  & $ -0.1$ & 15.698 0.022  & 15.577 0.010  &  15.046 0.009  &  14.882 0.010  &  14.664 0.009  &  LT  \\
 2009-10-15  &   55119.60  & $   0.2$ &   $-$               & 15.600 0.061  &  15.010 0.018  &  14.859 0.015  &  14.648 0.013  &  PROMPT \\
 2009-10-16  &   55121.45  & $   2.1$ & 15.740 0.040  & 15.581 0.017  &  15.045 0.013  &  14.861 0.013  &  14.639 0.014  &  A1.82 \\
 2009-10-18  &   55122.53  & $   3.1$ &   $-$               & 15.620 0.029  &  15.034 0.010  &  14.850 0.011  &  14.584 0.013  &  PROMPT \\
 2009-10-18  &   55123.37  & $   4.0$ & 15.858 0.040  & 15.623 0.019  &  15.046 0.015  &  14.812 0.005  &  14.563 0.006  &  A1.82 \\
 2009-10-20  &   55124.62  & $   5.2$ & 16.046 0.007  & 15.698 0.018  &  15.061 0.012  &  14.838 0.011  &  14.560 0.008  &  NTT \\
 2009-10-21  &   55125.52  & $   6.1$ &   $-$               & 15.771 0.015  &  15.034 0.027  &  14.885 0.023  &  14.615 0.020  &  PROMPT \\
 2009-10-23  &   55128.34  & $   8.9$ & 16.468 0.015  & 15.937 0.015  &  15.157 0.016  &  14.851 0.017  &  14.614 0.016  &  LT \\
 2009-10-24  &   55128.57  & $   9.2$ &   $-$               & 15.945 0.020  &  15.163 0.014  &  14.856 0.010  &  14.657 0.010  &  PROMPT \\
 2009-10-25  &   55129.54  & $ 10.1$ &   $-$               & 16.022 0.017  &    $-$               &    $-$               &    $-$               &  PROMPT \\
 2009-10-27  &   55131.54  & $ 12.1$ &   $-$               & 16.171 0.022  &  15.271 0.012  &  14.950 0.008  &  14.673 0.011  &  PROMPT \\
 2009-10-27  &   55132.60  & $ 13.2$ &   $-$               &    $-$              &  15.381 0.086  &      $-$             &      $-$             &  FORS2 \\
 2009-10-29  &   55133.52  & $ 14.1$ &   $-$               & 16.440 0.039  &  15.447 0.015  &  14.999 0.037  &  14.759 0.052  &  PROMPT  \\
 2009-11-02  &   55137.52  & $ 18.1$ &   $-$               & 16.860 0.031  &  15.697 0.017  &  15.199 0.016  &  14.866 0.016  &  PROMPT  \\
 2009-11-03  &   55138.54  & $ 19.1$ &   $-$               & 16.929 0.045  &  15.726 0.089  &  15.258 0.019  &  14.924 0.013  &  PROMPT \\
 2009-11-03  &   55139.33  & $ 19.9$ & 17.632 0.103  & 16.937 0.012  &  15.797 0.010  &  15.235 0.008  &  14.906 0.010  &   LT \\
 2009-11-04  &   55140.38  & $ 21.0$ & 17.600 0.035  & 17.043 0.014  &  15.874 0.018  &  15.322 0.029  &  14.969 0.019  &   NOT \\
 2009-11-06  &   55141.52  & $ 22.1$ &   $-$               & 17.081 0.040  &    $-$               &    $-$               &    $-$               &  PROMPT \\
 2009-11-09  &   55144.56  & $ 25.2$ &   $-$               &   $-$               &    $-$               &  15.548 0.023  &  15.120 0.024  &  PROMPT \\
 2009-11-11  &   55147.33  & $ 27.9$ & 17.969 0.045  & 17.381 0.014  &  16.215 0.008  &  15.586 0.009  &  15.258 0.011  &   LT \\
 2009-11-12  &   55147.54  & $ 28.1$ &   $-$               & 17.427 0.045  &  16.220 0.021  &  15.666 0.017  &  15.247 0.014  &  PROMPT  \\
 2009-11-12  &   55147.54  & $ 28.1$ &   $-$               &   $-$               &  16.235 0.018  &    $-$               &    $-$               &  PROMPT  \\
 2009-11-13  &   55149.34  & $ 29.9$ & 18.070 0.070  & 17.440 0.014  &  16.274 0.009  &  15.652 0.007  &  15.340 0.009  &   LT        \\
 2009-11-13  &   55149.37  & $ 30.0$ & 18.050 0.041  & 17.437 0.014  &  16.286 0.009  &  15.780 0.012  &  15.330 0.013  &   NOT  \\
 2009-11-14  &   55149.53  & $ 30.1$ &   $-$               &   $-$               &  16.305 0.029  &  15.728 0.018  &  15.303 0.015  &  PROMPT \\
 2009-11-18  &   55154.41  & $ 35.0$ & 18.080 0.192  & 17.584 0.017  &  16.482 0.015  &  15.906 0.012  &  15.397 0.029  &   CA \\
 2009-11-19  &   55154.54  & $ 35.1$ &   $-$               & 17.605 0.054  &  16.502 0.049  &  15.936 0.066  &  15.433 0.056  &  PROMPT \\
 2009-11-22  &   55157.52  & $ 38.1$ & 18.249 0.101  & 17.683 0.032  &  16.558 0.029  &  15.980 0.006  &  15.511 0.017  &   NTT \\
 2009-11-23  &   55158.26  & $ 38.9$ & 18.141 0.003  & 17.702 0.015  &  16.568 0.014  &  16.000 0.013  &  15.508 0.012  &   CA \\
 2009-11-23  &   55159.35  & $ 39.9$ & 18.250 0.104  & 17.718 0.048  &  16.564 0.050  &  15.982 0.012  &  15.594 0.010  &   LT  \\
 2009-11-24  &   55159.56  & $ 40.2$ &  $-$                & 17.721 0.25    &  16.62   0.20    &  16.014 0.017  &  15.576 0.016  &  PROMPT \\
 2009-11-27  &   55162.53  & $ 43.1$ &  $-$                & 17.800 0.077  &  16.680 0.057  &  16.047 0.018  &  15.633 0.015  &  PROMPT   \\
 2009-12-01  &   55167.32  & $ 47.9$ &  18.333 0.098  & 17.830 0.055 &  16.751 0.059 &  16.146 0.012  &  15.793 0.012  &   LT \\
 2009-12-06  &   55172.34  & $ 52.9$ & 18.336 0.068  & 17.84  0.11     &  16.80   0.11    &  16.277 0.009  &  15.869 0.008  &   LT \\
 2009-12-28  &   55194.30  & $74.9 $ &     $-$             &     $-$             &      $-$             &  16.664 0.075  &  16.196 0.065  &   CA \\
 2009-12-29  &   55195.29  & $75.9 $ & 18.456 0.172  & 18.09   0.20    &  17.14   0.19    &      $-$             &  16.214 0.045  &   CA \\
 2010-01-02  &   55199.27  & $79.9$  & $>$ 17.97      &     $-$             &  17.229 0.058  &  16.746 0.061  &  16.273  0.060  &   CA \\
 2010-01-07  &   55204.36  & $85.0$  & 18.464 0.067  & 18.143 0.012  &  17.324 0.010  &  16.869 0.016  &  16.412 0.015  &   NOT \\
 2010-06-18  &   55368.90  & $249.5$&     $-$             & 20.483 0.088  &  20.164 0.081  &  19.368 0.042  &  19.273 0.051  &   NTT \\
 2010-07-14  &   55392.57  & $273.2$&     $-$             & 21.44  0.57     &  20.68   0.42    &  19.81    0.16   &  19.97  0.15     &   CA \\
 2010-10-04  &   55473.78  & $354.4$&     $-$             &     $-$             &  22.00    0.20   &    21.39 0.16    &       $-$            &   NTT \\
 2010-10-11  &   55480.54  & $361.1$&     $-$             &     $-$             &    $-$               &  21.30   0.43    &       $-$            &   FORS2   \\
\hline
 \end{tabular}\\
$^a$ The errors are computed taking into account 
both the uncertainty of the PSF fitting of the SN magnitude and 
the uncertainty due to the background contamination 
(computed by an artificial-star experiment).
$^b$Relative to the $B$-band maximum (JD = 2,455,119.4). 
$^c~$ IAUC =  IAU Circular 8232.
$~$ A1.82 $=$ Asiago 1.82~m telescope and AFOSC; pixel scale $=$ 0.473 arcsec pixel$^{-1}$. 
$~$ NOT $=$ Nordic Optical Telescope and ALFOSC; pixel scale $=$ 0.188 arcsec pixel$^{-1}$. 
$~$ TNG $=$ Telescopio Nazionale Galileo and DOLORES; pixel scale $=$ 0.25 arcsec pixel$^{-1}$. 
$~$ PROMPT $=$ PROMPT telescopes and CCD camera Alta U47UV E2V CCD47-10; pixel scale $=$ 0.590 arcsec pixel$^{-1}$.
$~$ FORS2 = 8.2~m ESO-VLT Telescope and FORS2,  pixel scale $=$ 0.124 arcsec pixel$^{-1}$.
$~$ CA = Calar Alto 2.2m Telescope + CAFOS;  pixel scale $=$ 0.53 arcsec pixel$^{-1}$.
$~$ LT $=$ Liverpool Telescope and RATCam; pixel scale $=$ 0.279 arcsec pixel$^{-1}$.
$~$ NTT $=$ New Technology telescope and EFOSC2; pixel scale $=$ 0.24 arcsec pixel$^{-1}$. 
$~$ NAR $=$ Newtonian Astrograph Reflector and Canon EOS 350D camera.
$~$ AN1 $=$ AlbaNova 1~m telescope and DV438.
$~$ RCOS20 $=$ 20$''$ RCOS and STL11K; pixel scale $=$ 0.9 arcsec pixel$^{-1}$.
\end{minipage}
\end{table*}

\begin{table*}
 \centering
 \begin{minipage}{160mm}
  \caption{Optical photometry of SN 2009jf (AB magnitudes in Sloan system)$^a$.}
  \label{tabsloan}
  \begin{tabular}{@{}ccccccccc@{}}
  \hline
  Date & JD $-$ & Phase$^{b}$& $u$ & $g$ & $r$ & $i$ & $z$ & Source$^{c}$\\
        & 2,400,000\\
\hline
 2009-10-04  &  55109.41 & $-10.0 $&                $-$                &   $-$                 &  15.767 0.013  &  15.808 0.013  &  $-$                 &  LT \\   
 2009-10-06  &  55111.38 &$  -8.0 $ &                $-$                &   $-$                 &  15.518 0.008  &  15.572 0.011  &  $-$                 &  LT  \\
 2009-10-08  &  55113.39 & $ -6.0 $ &                16.688 0.018 &   15.569 0.008  &  15.343 0.009  &  15.388 0.009  &  15.438 0.016  &  LT  \\
 2009-10-08  &  55113.55 & $ -5.8 $ &                 $-$               &   15.428 0.092  &  15.304 0.036  &  15.309 0.040  &  15.334 0.077  &  PROMPT \\
 2009-10-10  &  55115.39 & $ -4.0 $ &                16.644 0.022 &   15.394 0.017  &  15.208 0.010  &  15.261 0.008  &  15.240 0.016  &  LT  \\  
 2009-10-12  &  55116.56 & $ -2.8 $ &                $-$                &   15.390 0.015  &  15.114 0.015  &  15.185 0.015  &  15.207 0.017  &  PROMPT \\
 2009-10-12  &  55117.41 & $ -2.0 $ &                16.552 0.024 &   15.327 0.015  &  15.098 0.009  &  15.138 0.010  &  15.169 0.017  &  LT    \\
 2009-10-13  &  55117.62 & $ -1.8 $ &                $-$                &   15.350 0.024  &  15.129 0.028  &  15.150 0.019  &  15.090 0.028  &  PROMPT \\
 2009-10-14  &  55119.35 & $ -0.1$  &                16.568 0.024 &   15.328 0.018  &  15.040 0.010  &  15.056 0.009  &  15.080 0.012  &  LT  \\
 2009-10-15  &  55119.62 & $   0.2$  &                $-$                &   15.305 0.024  &  15.009 0.011  &  15.005 0.019  &  15.045 0.021  &  PROMPT \\
 2009-10-18  &  55122.53 & $  3.1$   &                $-$                &   $-$                 &  14.963 0.013  &  15.002 0.016  &  14.942 0.024  &  PROMPT \\
 2009-10-21  &  55125.53 & $  6.1$   &                $-$                &   15.429 0.012  &  14.944 0.018  &  14.987 0.016  &  14.929 0.023  &  PROMPT \\
 2009-10-23  &  55128.34 & $  8.9$   &                17.301 0.021 &   15.527 0.015  &  15.020 0.017  &  15.010 0.016  &  14.978 0.010  &  LT \\
 2009-10-24  &  55128.51 & $  9.1$   &                $-$                &   15.557 0.011  &  $-$                 &  $-$                 &  $-$                 &  PROMPT \\
 2009-10-25  &  55129.55 & $10.1 $  &                $-$                &   15.575 0.011  &  $-$                 &  $-$                 &  $-$                 &  PROMPT \\
 2009-10-27  &  55131.52 & $12.1 $  &                $-$                &   15.711 0.010  &  15.112 0.012  &  15.064 0.010  &  15.030 0.014  &  PROMPT \\
 2009-10-28  &  55132.53 & $13.1 $  &                $-$                &   15.777 0.016  &  15.178 0.036  &  15.125 0.016  &  15.065 0.014  &  PROMPT \\
 2009-11-03  &  55138.54 & $19.1 $  &                $-$                &   16.314 0.061  &  15.456 0.016  &  15.353 0.010  &  15.244 0.037  &  PROMPT \\
 2009-11-03  &  55139.34 & $19.9 $  &                18.370 0.039 &   16.408 0.008  &  15.499 0.010  &  15.336 0.010  &  15.263 0.016  &  LT    \\
 2009-11-05  &  55140.52 & $21.1 $  &                $-$                &   16.400 0.012  &  $-$                 &  15.441 0.012  &  15.288 0.030  &  PROMPT \\
 2009-11-11  &  55147.33 & $27.9 $  &                $-$                &   $-$                 &  15.869 0.016  &  15.710 0.006  &  15.479 0.024  &  LT \\
 2009-11-13  &  55148.78 & $29.4 $  &                $-$                &   16.967 0.017  &  15.969 0.015  &  15.804 0.006  &  15.499 0.015  &  PROMPT \\
 2009-11-13  &  55149.35 & $29.9 $  &                18.966 0.136 &   17.004 0.018  &  15.961 0.015  &  15.789 0.015  &  15.531 0.012  &  LT \\
 2009-11-19  &  55154.52 & $35.12$ &                $-$                &   $-$                 &  16.128 0.017  &  15.990 0.010  &  15.605 0.008  &  PROMPT \\
 2009-11-23  &  55159.36 & $40.0 $  &                19.122 0.039 &   17.167 0.018  &  16.241 0.015  &  16.055 0.009  &  15.640 0.009  &  LT \\
 2009-11-24  &  55159.52 & $40.1 $  &                $-$                &    $-$                &  16.199 0.017  &  16.081 0.019  &  15.647 0.019  &  PROMPT \\
 2009-12-01  &  55167.32 & $47.9 $  &                19.151 0.098 &   17.249 0.026  &  16.381 0.012  &  16.240 0.010  &  15.785 0.013  &  LT \\    
 2009-12-02  &  55167.55 & $48.1 $  &                $-$                &   $-$                 &  16.396 0.013  &  16.259 0.016  &  $-$                 &  PROMPT \\
 2009-12-06  &  55172.34 & $ 52.3$  &                19.206 0.056 &   17.283 0.017  &  16.429 0.013  &  16.301 0.012  &  15.892 0.044  &  LT \\
\hline	 		 
\end{tabular}\\
$^a$ The errors are computed taking into account 
both the uncertainty of the PSF fitting of the SN magnitude and 
the uncertainty due to the background contamination 
(computed by an artificial-star experiment).
$^b$Relative to the $B$-band maximum (JD = 2,455,119.4).
$^c$ PROMPT $=$ Prompt telescopes and CCD camera Alta U47UV E2V CCD47-10; pixel scale $=$ 0.590 arcsec pixel$^{-1}$.
$~$ LT $=$ Liverpool Telescope and RATCam; pixel scale $=$ 0.279 arcsec pixel$^{-1}$.
\end{minipage}
\end{table*}

\begin{table*}
  \caption{Infrared photometry of SN 2009jf$^a$.}
  \label{tabinf}
  \begin{tabular}{@{}ccccccc@{}}
    \hline
  Date & JD $-$ & Phase$^{b}$& $J$ & $H$ & $K$ & Source$^{c}$\\
        & 2,400,000\\
\hline
 02-10-2009 & 55106.59   & $-12.8 $ &                 15.565 0.029  &  15.511 0.042  &  15.356 0.052  &     NICS \\
 07-10-2009 & 55111.45   & $-7.9   $ &                 14.838 0.016  &  14.725 0.018  &  14.527 0.021  &     NICS \\
 13-10-2009 & 55117.31   & $-2.1   $ &                 14.42   0.11    &  14.393 0.093  &  14.29   0.14    &     NICS \\
 20-10-2009 & 55125.64   & $ 6.2    $ &                 14.285 0.011  &  14.151 0.011  &  13.956 0.015  &     SOFI  \\
 23-10-2009 & 55128.35   & $ 8.9    $ &                 14.305 0.067  &  14.335 0.089  &   $-$                &     LT     \\
 24-10-2009 & 55129.39   & $ 10.0  $ &                 14.32   0.13    &  14.19   0.15    &   $-$                &     LT     \\
 27-10-2009 & 55132.02   & $12.6   $ &                 14.36   0.20    &  14.18   0.20    &   $-$                &     REM  \\
 06-11-2009 & 55141.16   & $ 21.8  $ &                 14.52   0.14    &  14.40   0.11    &  14.30   0.11    &     REM  \\
 11-11-2009 & 55146.08   & $ 26.7  $ &                 14.77   0.38    &  14.61   0.41    &  14.637 0.096  &     REM  \\
 15-11-2009 & 55150.14   & $ 30.7  $ &                 14.81   0.47    &  14.79   0.44    &  14.70   0.26    &     REM  \\
 19-11-2009 & 55154.14   & $ 34.7  $ &                 14.99   0.37    &  14.89   0.37    &  $>$ 14.322    &      REM  \\ 
 23-11-2009 & 55158.52   & $ 39.1  $ &                 15.069 0.020  &  14.897 0.024  &  14.924 0.029  &      SOFI \\
 24-11-2009 & 55159.05   & $ 39.6  $ &                 15.11   0.25    &  14.817 0.25    &  15.34   0.23    &      REM  \\
 08-12-2009 & 55174.32   & $ 54.9  $ &                 15.600 0.056  &  15.140 0.052  &   $-$                &      LT    \\
 02-07-2010 & 55378.68   & $ 259.3$ &                 19.34   0.30    &   18.57  0.30    &  19.08   0.30    &      NICS  \\
\hline
\end{tabular}\\
$^a$ The errors are computed taking into account 
both the uncertainty of the PSF fitting of the SN magnitude and 
the uncertainty due to the background contamination 
(computed by an artificial-star experiment). $^b$ Relative to the $B$-band maximum (JD = 2,455,119.4). 
$^c~$ NICS  $=$ Telescopio Nazionale Galileo and NICS; pixel scale $=$ 0.25 arcsec pixel$^{-1}$. 
$~$ LT $=$ Liverpool Telescope and SupIRCam near-IR camera; pixel scale $=$ 0.41 arcsec pixel$^{-1}$.
$~$ REM $=$  Rapid Eye Mount 60 cm telescope and REMIR ; pixel scale $=$ 1.2 arcsec pixel$^{-1}$.
$~$ SOFI $=$  NTT telescope and SOFI; pixel scale $=$ 0.29 arcsec pixel$^{-1}$.
\end{table*}

\begin{table*}
  \caption{\emph{Swift} photometry of SN 2009jf$^a$.}
  \label{tabswift}
  \begin{tabular}{@{}cccccccc@{}}
    \hline
 Date & JD $-$ & Phase$^{b}$& $uvw1$ & $U$ &  $B$ &  $V$ & Source\\
        & 2,400,000\\
\hline
 2009-09-30 & 55104.68  & $-14.7 $ &        18.03  0.07  &  18.01  0.19  &  17.31  0.05  &   16.80  0.07   &  SWIFT  \\
 2009-10-01 & 55105.55  & $-13.8 $ &        18.02  0.07  &  17.63  0.17  &  17.13  0.05  &   16.53  0.07   &  SWIFT  \\
 2009-10-04 & 55108.97  & $-10.4 $ &        17.90  0.06  &  16.47  0.14  &  16.28  0.03  &   15.80  0.05   &  SWIFT  \\
 2009-10-06 & 55111.31  & $-8.1  $  &        18.04  0.07  &  16.03  0.09  &  15.90  0.03  &   15.47  0.04   &  SWIFT  \\
 2009-10-08 & 55113.26  & $-6.1  $  &        18.02  0.08  &  15.88  0.06  &  15.77  0.03  &   15.35  0.04   &  SWIFT  \\
 2009-10-10 & 55115.14  & $-4.3  $  &        17.85  0.07  &  15.74  0.06  &  15.67  0.03  &   15.22  0.04   &  SWIFT  \\
 2009-10-12 & 55117.21  & $-2.2  $  &        18.02  0.08  &  15.76  0.08  &  15.60  0.04  &   15.09  0.04   &  SWIFT  \\
 2009-10-14 & 55119.15  & $-0.2  $  &        18.00  0.06  &  15.73  0.05  &  15.60  0.04  &   15.03  0.04   &  SWIFT  \\ 
 2009-10-16 & 55120.54  & $ 1.1   $  &        17.97  0.07  &  15.82  0.10  &  15.62  0.03  &   15.03  0.04   &  SWIFT  \\
 2009-10-18 & 55122.82  & $ 3.4   $  &        17.92  0.06  &  15.98  0.08  &  15.62  0.03  &   15.06  0.04   &  SWIFT  \\ 
 2009-10-20 & 55124.55  & $ 5.1   $  &        18.10  0.07  &  16.09  0.07  &  15.73  0.03  &   15.07  0.04   &  SWIFT  \\
 2009-11-01 & 55137.14  & $ 17.7 $  &        18.04  0.07  &  17.32  0.07  &  16.75  0.04  &   15.65  0.04   &  SWIFT  \\ 
 2009-11-04 & 55139.56  & $ 20.2 $  &        18.02  0.08  &  17.68  0.08  & 16.915 0.04  &   15.80  0.05   &  SWIFT  \\
 2009-11-07 & 55142.64  & $ 23.2 $  &           $-$           &  17.72  0.13  &  17.21  0.06  &    $-$              &  SWIFT  \\
\hline
\end{tabular}\\
$^a$ The $U$-band data have been reduced using the template subtraction 
technique within the QUBA pipeline. The S-correction has been applied.  
The errors have been computed taking into account both the uncertainty of the PSF fitting of the SN magnitude and  the uncertainty due to the background contamination (computed by an artificial-star experiment). 
The $uvw1$, $B$ and $V$ bands have been reduced using the SWIFT pipeline. The $uvm2$ and $uvw2$ 
bands have been measured with the SWIFT pipeline but are not reported as they are strongly contaminated by the nearby cluster.
$^b$Relative to the $B$-band maximum (JD = 2,455,119.4).\\
\end{table*}

\end{document}